\documentclass[oldversion]{aa} 
\usepackage{graphicx}
\usepackage{longtable}
\usepackage{txfonts}
\usepackage{rotating}
\usepackage{lscape}
\newcommand{\gsim}{\;\lower.6ex\hbox{$\sim$}\kern-7.75pt\raise.65ex\hbox{$>$}\;}
\newcommand{\lsim}{\;\lower.6ex\hbox{$\sim$}\kern-7.75pt\raise.65ex\hbox{$<$}\;}

\begin{document}
\title{The extreme chemistry of multiple stellar populations in the metal-poor globular 
cluster NGC~4833\thanks{Based on observations collected at 
ESO telescopes under programmes 083.D-0208 and 68.D-0265}\fnmsep\thanks{
   Tables 2, 6, 7, 8, 9, and 11 are only available in electronic form at the CDS via anonymous
   ftp to {\tt cdsarc.u-strasbg.fr} (130.79.128.5) or via
   {\tt http://cdsweb.u-strasbg.fr/cgi-bin/qcat?J/A+A/???/???}}
 }

\author{
E. Carretta\inst{1},
A. Bragaglia\inst{1},
R.G. Gratton\inst{2},
V. D'Orazi\inst{3,4},
S. Lucatello\inst{2},
Y. Momany\inst{2,5},
A. Sollima\inst{1},
M. Bellazzini\inst{1},
G. Catanzaro\inst{6}
\and
F. Leone\inst{7}
}

\authorrunning{E. Carretta et al.}
\titlerunning{Multiple stellar populations in NGC~4833}

\offprints{E. Carretta, eugenio.carretta@oabo.inaf.it}

\institute{
INAF-Osservatorio Astronomico di Bologna, Via Ranzani 1, I-40127 Bologna, Italy
\and
INAF-Osservatorio Astronomico di Padova, Vicolo dell'Osservatorio 5, I-35122
 Padova, Italy
\and
Dept. of Physics and Astronomy, Macquarie University, Sydney, NSW, 2109 Australia 
\and
Monash Centre for Astrophysics, Monash University, School of Mathematical 
Sciences, Building 28, Clayton VIC 3800, Melbourne, Australia
\and
European Southern Observatory, Alonso de Cordova 3107, Vitacura, Santiago, Chile
\and
INAF-Osservatorio Astrofisico di Catania, Via S.Sofia 78, I-95123 Catania, Italy
\and
Dipartimento di Fisica e Astronomia, Universit\`a di Catania, Via S.Sofia 78, 
 I-95123 Catania, Italy 
  }

\date{}

\abstract{Our FLAMES survey of Na-O anticorrelation in globular clusters
(GCs) is extended to NGC 4833, a metal-poor GC with a long blue tail on the
horizontal branch (HB). We present the abundance analysis for a large sample of
78 red giants based on UVES and GIRAFFE spectra acquired at the ESO-VLT. We
derived abundances of Na, O, Mg, Al, Si, Ca, Sc, Ti, V, Cr, Mn, Fe, Co, Ni, Cu,
Zn, Y, Ba, La, Nd. This is the first extensive study of this cluster from high
resolution spectroscopy. On the scale of our survey, the metallicity of NGC~4833
is  [Fe/H]$=-2.015 \pm0.004 \pm0.084$ dex ($rms=0.014$ dex) from 12 stars
observed with UVES, where the first error is from statistics and the second one
refers to the systematic effects. The iron abundance in NGC~4833 is homogeneous
at better than 6\%. On the other hand, the light elements involved in
proton-capture reactions at high temperature show the large star-to-star
variations observed in almost all GCs studied so far. The Na-O anticorrelation
in NGC~4833 is quite extended, as expected from the high temperatures reached by
stars on the HB, and NGC~4833 contains a conspicuous fraction of stars with
extreme [O/Na] ratios. More striking is the finding that large star-to-star
variations are seen also for Mg, which spans a range of more than 0.5 dex in
this GC. Depletions in Mg are correlated to the abundances of O and
anti-correlated with Na, Al, and Si abundances. This pattern suggests the action
of nuclear processing at unusually high temperatures, producing the extreme
chemistry observed in the stellar generations of NGC~4833. This extreme changes
are also seen in giants of the much more massive GCs M~54 and $\omega$ Cen, and
our conclusion is that NGC~4833 has probably lost a conpicuous fraction of its
original mass due to bulge shocking, as also indicated by its orbit.}

\maketitle

\section{Introduction}

Globular clusters (GCs) are the brightest relics of the early phases of galaxy
formation.  Their study provides basic information on the early evolution of
their host galaxy.  In the last few years, it has become apparent that the
episodes leading to the  formation of these massive objects were complex. The
main evidence for this complexity comes from the presence of chemical
inhomogeneities that in most cases are limited to  light elements involved in
H-burning at high temperature (He, CNO, Na, Mg, Al), though  in a few other GCs
also star-to-star variations of heavier elements are found (see reviews  by
Gratton et al. 2004; and Gratton et al. 2012). The extensive spectroscopic
survey we  are conducting (see e.g. Carretta et al. 2009a,b) revealed that these
inhomogeneities are  ubiquitous among GCs, but their range changes
from cluster-to-cluster, being  driven mainly by the total mass of the GCs,
though also metallicity and possibly other parameters play a role. 

Similar results have been obtained by other authors, both using similar analysis
methods (e.g. Ramirez and Cohen 2002, 2003; Marino et al. 2008, Johnson and
Pilachowski 2012) and photometric ones (e.g. Milone et al. 2012, 2013 and
references therein). This led to a general scenario where GCs host multiple
stellar populations, often considered to correspond to different generations,
where the younger stars (second generation stars) are formed mainly  or
even exclusively from the ejecta of older ones (first or primordial generation
stars: see e.g. D'Ercole et al. 2008, Decressin et al. 2008), though
alternatives are also being considered (see e.g. Bastian et al. 2013a). 

There are several key aspects that still require understanding. It is important
to establish which of the first generation stars polluted the
material from which second generation stars formed. This is relevant to
set both the involved time-scale and  the mass budget. Given that no variation
in Fe abundance is observed in most GCs (see e.g. Carretta et al.
2009c), supernovae should play at most a very minor role, apart from a few
exceptional cases. This clearly makes the chemical evolution of GCs very
different from the one typically observed in galaxies, and requires a
restriction of the mass range of the polluters. Several  candidates have been
proposed: the most massive among intermediate-mass stars during the asymptotic
giant branch (AGB) phase (Ventura et al. 2001);
fast rotating massive stars (Decressin et al. 2007); massive  binaries (de Mink
et al. 2009); novae (Maccarone and Zurek 2012). 

Arguments in favour or against each of these candidates exist, and it is not
even clear if they should be considered as mutually exclusive. Scenarios based
on different polluters produce different expectations about some observables
(e.g. Valcarce and Catelan 2011).
For instance, those involving massive stars are characterized by a very short
time-scale, which is possibly supported by the lack of  evidence for multiple
generation of stars in present-day massive clusters (Bastian et al. 2013b),
though it is not well clear that such objects are as massive as the Milky Way 
GCs were when they formed. On the other hand, they have difficulties in
producing a well defined threshold in cluster mass for the phenomenon (see
Carretta et al. 2010a) as well as clearly separated stellar populations, as
observed in several typical clusters from both photometry and spectroscopy
(NGC~2808: D'Antona et al. 2005, Carretta et al. 2006, Piotto et al. 2007; M~4:
Marino et al. 2008, Carretta et al. 2009a,b; NGC~6752: Carretta et al. 2012a,
Milone et al. 2013; 47~Tuc: Milone et al. 2012, Carretta et al. 2012b, to quote
a few examples).

Providing high quality data on more GCs is clearly needed to strengthen the
results obtained so far. Spectroscopic analysis of rather large samples of stars
may for instance be used to show if stars can be divided into discrete groups in
chemical composition, that may more easily be explained as evidence for multiple
episodes of star formation, or rather distribute continuously. Also, very
stringent limits on star-to-star variations in Fe abundances  as observed in
many GCs by Carretta et al. (2009c) or Yong et al. (2013) severely limit
the possibility that SNe contributed to chemical evolution, which is one of the
basic problems to be faced by short-timescale scenarios of cluster formation.

Following our earlier work (e.g. Carretta et al. 2006, 2009a,b), we present here
the results of the analysis of large samples of spectra in NGC~4833, focussing
on the Na-O anticorrelation, but also providing data for Fe, Mg, Al, Si, Ca, Ti,
and other elements. This is the first extensive analysis of high dispersion
spectra for this cluster since only two red giant branch (RGB) stars were
analysed by Gratton and Ortolani (1989) and one by Minniti et al. (1996), which
of course prevented any discussion on chemical inhomogeneities. NGC~4833
($M_V=-8.16$: Harris 1996) has been classified as an old halo cluster with a
moderately extended blue horizontal branch (Mackey and van den Bergh 2005) or an
inner halo cluster (Carretta et al. 2010a). Studies of variable stars have been
presented by Murphy and Darragh (2012, 2013), who found 17 RR Lyrae and six SX
Phoenicis stars. The mean period of the RR Lyrae allows to classify the cluster
as Oosterhoff II. NGC~4833 is seen projected close to the Coal Sack, and has a
moderately large reddening ($E(B-V)=0.32$) with not negligible variations across
the cluster. Casetti-Dinescu et al. (2007) determined a very eccentric orbit
($e=0.84$) that brings the cluster very close to the Galactic center and never
far from the Galactic plane ($z_{\rm max}=1.8$~kpc, $r_a=7.7$~kpc,
$r_p=0.7$~kpc). Orbital parameters are similar to those of NGC~5986, so that a
tentative association between the two clusters was  proposed by Casetti-Dinescu
et al.

The structure of this paper is as follows. Observations, radial velocities and
kinematics are presented in Section 2, while Section 3 is devoted to the
abundance analysis, whose results are illustrated in Section 4. Our findings are
discussed in Section 5 and summarised in Section 6.

\begin{figure}
\centering
\includegraphics[scale=0.44]{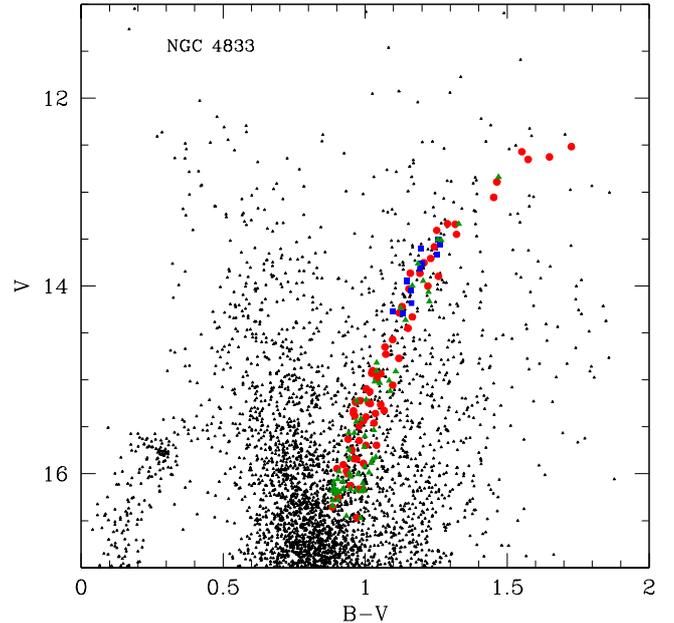}
\caption{The $V,B-V$ CMD of NGC~4833 (open circles). Stars selected for the
present study are plotted as filled, larger symbols: blue squares are the stars 
observed with UVES, red circles are stars with GIRAFFE spectra, and green
triangles are stars observed with GIRAFFE but not analyzed because found to be
non members from their RV (see text).}
\label{f:cmdsel4833}
\end{figure}

\section{Observations}

The photometric catalog for NGC~4833 is based on $UBVI$ data (ESO programme 
68.D-0265, PI Ortolani), collected at the Wide-Field Imager at the 2.2-m ESO-MPI
telescope   on 17-21 February 2002. The WFI covers a total field of view of
$34^{\prime}\times~33^{\prime}$, consisting of $8$, $2048\times 4096$ EEV-CCDs
with a pixel size of $0\farcs238$. The collected images were dithered in order
to cover the gaps between the CCDs, and the exposure times were divided into
shallow and deep so as not to saturate the bright red giant stars and, at the
same time, sample the faint main sequence stars with a good signal to noise.

The de-biasing and flat-fielding reduction of the CCD mosaic raw images employed
the IRAF package MSCRED (Valdes 1998), while the stellar photometry was derived
via the use of the DAOPHOT and ALLFRAME programs (Stetson 1994). For specific
details of the photometric reduction process see Momany et al. (2004).

The absolute photometric calibration primarily employed the use of $UBVI$
standard stars from Landolt (1992), as well as secondary stars from the Stetson
(http://cadcwww.hia.nrc.ca/standards/) library, which provides more numerous and
relatively fainter $BVI$ standards. The uncertainties in the absolute flux
calibration were of the order of 0.06, 0.03, 0.03, and 0.04 mag, for $UBVI$,
respectively. The NGC~4833 catalog was not corrected for sky-concentration,
caused by the spurious reflections of light and its subsequent redistribution in
the focal plane (Manfroid et al. 2001). As a consequence, photometric
comparisons to independent catalogs (based on smaller field of view data) might
reveal systematic offsets as a function of the distance from the center of the
WFI mosaic. 

We selected a pool of stars lying near the RGB ridge line in
the colour magnitude diagram (CMD) and without close neighbours, i.e. without
any star closer than 3 arcsec (we included also cases with neighbours between 2
and 3 arcsec, but only if fainter by more than 2 mag). The {\sc FPOSS} tool was
used to allocate the FLAMES (Pasquini et al. 2002) fibres. 

The stars in our spectroscopic sample are indicated as coloured, filled symbols
in Fig.~\ref{f:cmdsel4833}. We clearly see that NGC~4833 is heavily contaminated
by field stars of the Galactic disk and bulge, given its present location at
$l=303.60\deg, ~b=-8.02\deg$. Even if the reddening toward the cluster is 
large and differential (Melbourne\& Guhathakurta 2004), thus complicating the
analysis, the excision of non-members via radial velocity and the use of
infrared filters for temperature determinations as done in similar cases (see
e.g. Gratton et al. 2006, 2007), make it feasible.

The log of the observations is given in Table~\ref{t:logobs}. We obtained two
exposures with the high resolution GIRAFFE grating and the setup HR11 
covering the Na~{\sc i}
5682-88~\AA\ doublet and two exposures with the setup HR13 including the
[O~{\sc i}] forbidden lines at 6300-63~\AA.  Excluding a non-member star and
another not useful due to the low S/N of the spectrum, we observed a total of 12
(bright) giants with the fibers feeding the UVES spectrograph (Red Arm, with
spectral range from 4800 to 6800~\AA\ and R=47,000, indicated as blue squares in
Fig.~\ref{f:cmdsel4833}) and 73 with GIRAFFE (seven are in common). 
The median values of the S/N ratio of spectra obtained with UVES and with 
the HR11 and HR13 setups of GIRAFFE are 90, 95, and 140, respectively.

\begin{figure}
\centering
\includegraphics[scale=0.44]{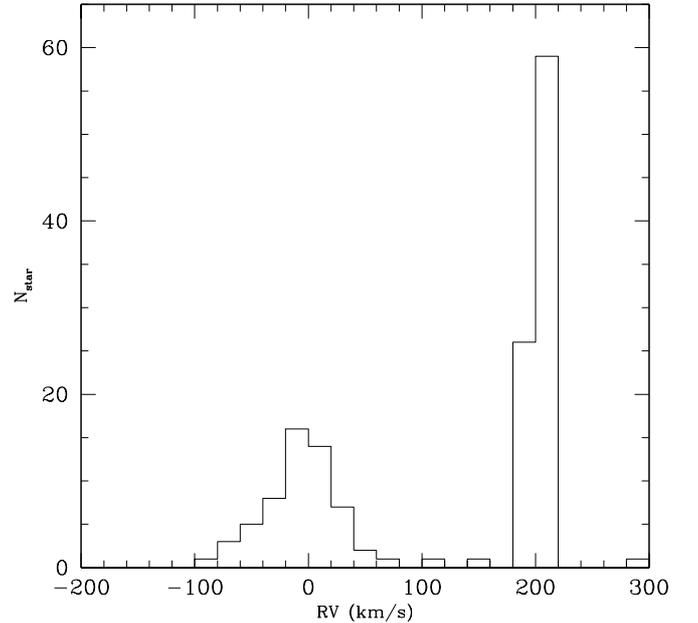}
\caption{Distribution of heliocentric radial velocities (RVs) for stars observed
in NGC~4833.}
\label{f:rvhel48}
\end{figure}

We used the 1-D, wavelength calibrated spectra as reduced by the ESO personnel 
with the dedicated FLAMES pipelines. 
Radial velocities (RV) for stars observed with the GIRAFFE spectrograph were
obtained using the  {\sc IRAF}\footnote{IRAF is  distributed by the National
Optical Astronomical Observatory, which are operated by the Association of
Universities for Research in Astronomy, under contract with the National Science
Foundation } task  {\sc FXCORR}, with appropriate templates, while those of the
stars observed with UVES were derived with the {\sc IRAF} task {\sc RVIDLINES}. 

The large RV of NGC~4833 makes it easy to isolate cluster stars from field
interlopers. In Fig.~\ref{f:rvhel48} we show the histogram of the heliocentric
RVs derived for all stars observed; the cluster is easily spotted as a narrow
and isolated peak around $V_r\simeq 200$~km~s$^{-1}$. We select as candidate
cluster members the 78 stars with $180.0\le RV\le$ 220~km~s$^{-1}$;
their membership is fully confirmed by the following chemical analysis, since
they all have the same iron abundance, within the uncertainties. The nearest
non-member in the velocity space has $RV=141.6\pm 0.7$~km~s$^{-1}$, $\sim
60$~km~s$^{-1}$ apart from the systemic velocity of the cluster.  A brief
overview of the cluster kinematics is provided in Sect.~\ref{kin}.

Our optical $B, V$ photometric data were integrated with $K$ band magnitudes 
from the Point Source Catalogue of 2MASS (Skrutskie et al. 2006) to derive
atmospheric parameters as described below, in Section 3.

Coordinates, magnitudes and heliocentric RVs are shown in Table~\ref{t:coo48}
(the full table is only available in electronic form at CDS).

\begin{figure}
\centering
\includegraphics[scale=0.44]{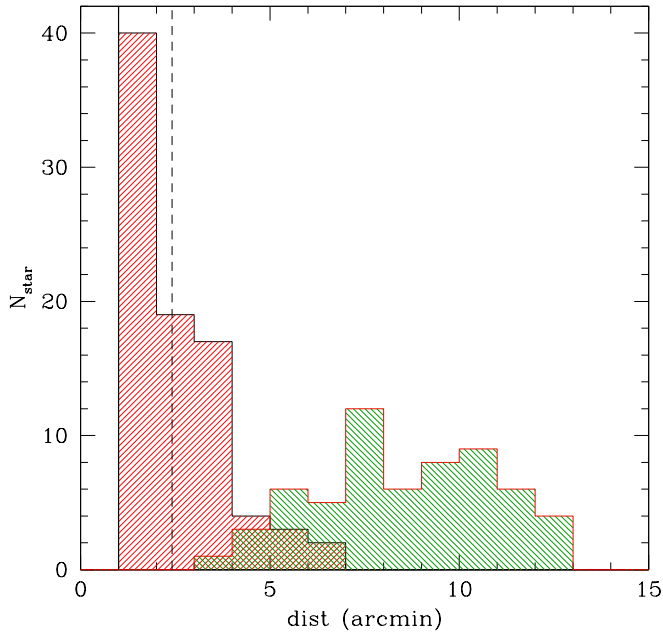}
\caption{Distribution of observed stars as a function of the radial distance
from the centre of NGC~4833. Member stars are indicated in red, while green is
used for non members. The solid heavy line indicates the core radius 
($r_c=1$ arcmin), and the dashed line the half-mass radius ($r_h=2.41$ arcmin) 
from Harris (1996).}
\label{f:hdist48}
\end{figure}

\begin{table}
\centering
\caption{Log of FLAMES observations for NGC 4833.}
\begin{tabular}{cccccc}
\hline\hline
   Date         &     UT       & exp. & grating & seeing & airmass\\
                &              & (s)  &         &  (")   &        \\ 
\hline
 Apr. 05, 2009  &02:11:37.395  &  2700   & HR11    & 1.07   & 1.599  \\
 Apr. 05, 2009  &02:58:02.303  &  2700   & HR11    & 1.09   & 1.519  \\
 Apr. 05, 2009  &03:54:25.968  &  2700   & HR13    & 0.64   & 1.462  \\
 Apr. 05, 2009  &04:40:47.286  &  2700   & HR13    & 0.59   & 1.446  \\
\hline
\end{tabular}
\label{t:logobs}
\end{table}

\subsection{Radial velocities and kinematics}
\label{kin}

There is no analysis in the literature dealing with the internal kinematics of
NGC~4833, and no estimate of  the central velocity dispersion $\sigma_0$. 
It is worth using our sample for a first investigation, albeit it is
not especially well suited for this purpose. Indeed our 78 genuine cluster
member stars are distributed (see Fig.~\ref{f:hdist48}) in the radial range
$1.0\le R/r_c\le 6.6$ or $0.4\le R/r_h\le 2.8$\footnote{The values for the core radius
$r_c$ and the half-light radius $r_h$ were taken, as all the cluster parameters
that will be used in the following, from the 2010 version of the Harris (1996)
catalog.}. Hence we cannot sample the kinematics in the cluster core and in the
outer halo (the tidal radius is $r_t= 17.8r_c= 7.4r_h$, since c=1.25). Our
analysis is fully homogenous with that performed in Bellazzini et al. (2012)
for the other clusters included in our survey: for further details,
we address the interested reader to that paper. The only difference is that here
mean velocity and velocity dispersions are estimated with the Maximum Likelihood
(ML) procedure described in Martin et al. (2007), that naturally takes into
account the effect of errors on the individual RV measures.

First of all we compared the RV estimates obtained from HR13 and HR11 spectra,
for the 45 stars that have valid RV measures from both setups. The mean
difference for the 35 stars brighter than V=15.5 (i.e. those with the smallest
individual errors) is 
$\langle RV_{13}-RV_{11}\rangle = -0.43 \pm 0.07$~km~s$^{-1}$. This shift was 
applied to the RV derived from HR11 spectra to
bring all the RV estimates to a common zero point. Then, for each star, we
adopted the RV estimated from HR13 spectra when present and when the individual
uncertainty is lower than the HR11 estimate, and the RV from HR11 in the other
cases.

Considering the whole sample of 78 stars we find an average velocity of 
$\langle RV \rangle=202.0\pm0.5$~km~s$^{-1}$ (and a dispersion of 
$4.1\pm0.3$~km~s$^{-1}$), in very good agreement with the value reported by 
Harris (1996), $\langle RV\rangle=200\pm1.2$~km~s$^{-1}$. Fitting a King (1966)
model with c=1.25 to the velocity dispersion curve we estimated the
central velocity dispersion $\sigma_0=5.0 \pm 1.0$~km~s$^{-1}$. We do not find
any statistically significant rotation: in the notation of Bellazzini et al.
(2012) the rotation
amplitude is $A_{rot}=0.9\pm 0.5$~km~s$^{-1}$ and $A_{rot}/\sigma_0=0.18\pm
0.11$. Hence NGC4833 fits excellently into the correlations between metallicity
(and horizontal branch morphology) and rotation found by Bellazzini et al., i.e.
metal-poor/blue HB clusters tend to have lower rotation than metal-rich/ red HB
clusters.

Using the King (1966) formula we estimate a dynamical mass
$M_{dyn}=1.3^{+0.6}_{-0.5}\times10^5~M_{\sun}$, and $(M/L)_V=0.84 \pm 0.45$, on
the low side of the range covered by Galactic GCs (see Sollima, Bellazzini \&
Lee 2012; Pryor \& Meylan 1993).

\section{Abundance analysis}

\subsection{Atmospheric parameters}

Following our tested procedure, effective temperatures $T_{\rm eff}$\ for our
targets were derived using an average relation between apparent magnitudes and
first-pass temperatures from $V-K$ colours and the calibrations of Alonso et al.
(1999, 2001). The rationale behind this adopted procedure is to decrease the
star to star errors in abundances due to uncertainties in temperatures, since
magnitudes are less affected by measure uncertainties than colours.
In the case of NGC~4833, affected by high and variable reddening we used the
apparent $K$ magnitudes in our relation with  $T_{\rm eff}$ because the impact
of the differential reddening on these magnitudes is very limited. This
procedure worked very well on other GCs heavily affected by large and
differential reddening patterns, like the bulge clusters  NGC~6441 (Gratton et
al. 2006, 2007) and NGC~6388 (Carretta et al. 2007a).
The adopted reddening $E(B-V)=0.32$ mag is from the Harris (1996)
catalogue, and it is in perfect agreement with the value derived in the accurate
study by Melbourne et al. (2000), who provided an associated error of 0.03 mag.
Using Table 3 of Cardelli et al. (1989) this translates into an error of 0.01
mag in $K$ magnitudes that, in turn, corresponds to an uncertainty in
temperature of 2.14 K, when coupled to the relation between effective
temperature and $K$ magnitude adopted for NGC~4833 in the present study.

Gravities were obtained from apparent magnitudes, assuming the $T_{\rm eff}$'s 
estimated above, bolometric corrections from Alonso et al. (1999), and the
distance modulus for NGC~4833 from Harris (1996).
We adopted a mass of 0.85~M$_\odot$\  for all stars and $M_{\rm bol,\odot} =
4.75$ as the bolometric magnitude for the Sun, as in our previous studies.

We eliminated trends in the relation between abundances from Fe~{\sc i} lines
and expected  line strength (Magain 1984) to obtain values of the microturbulent
velocity $v_t$.

Finally, using the above values we interpolated  within the Kurucz (1993) grid
of model atmospheres (with the option for overshooting on) to derive the final
abundances, adopting for each star the model with the appropriate atmospheric
parameters and whose abundances matched those derived from  Fe {\sc i} lines.
As discussed in Carretta et al. (2013a) this choise has a minimal impact on the
derived abundances with respect to models without overshooting.

\subsection{Elemental abundances}

Most of our derived abundances rest on analysis of equivalent widths ($EW$).
The code ROSA (Gratton 1988) was used as in previous papers to measure $EW$s
adopting a relationship between $EW$ and $FWHM$, as
described in detail in Bragaglia et al. (2001). Following the approach used in 
Carretta et al. (2007b) we first corrected the $EW$s from
GIRAFFE spectra to the system of the higher resolution UVES spectra, using seven
stars observed with both instruments. The correction has the form 
$EW_{\rm UVES} = 0.91(\pm0.02) \times EW_{\rm GIRAFFE} +1.06(\pm0.49)$ with an
$rms$ scatter of 5.4 m\AA\ and a Pearson correlation coefficient of $r=0.98$
from 121 lines.

Atomic parameters for the lines falling in the spectral range covered by UVES
spectra and by setups HR11 and HR13 in the GIRAFFE spectra are comprehensively
discussed in Gratton et al. (2003), together with the adopted solar reference
abundances.

Before coadding, each HR13 GIRAFFE spectrum was corrected for blending with
telluric lines due in particular to H$_2$O and O$_2$ near the forbidden 
[O~{\sc i}] line at 6300~\AA, using synthetic spectra as described in Carretta 
et al. (2006). Corrections for effects of departures from the LTE assumption
according to the descriptions by Gratton et al. (1999) were applied to the
derived Na abundances.

We derived abundances of O, Na, Mg, and Si among the elements participating to
the network of proton-capture reactions in H-burning at high temperature.
Additionally,  Al abundances  were obtained from the doublet Al~{\sc i}
6696-98~\AA\ for stars observed with UVES.

Beside Mg and Si (numbered among the proto-capture elements), we derived the
abundance of the $\alpha-$elements Ca and Ti~{\sc i}; Ti was obtained also from
the singly ionized species in stars with UVES spectra.
Abundances for the Fe-peak elements Sc~{\sc ii}, V~{\sc i}, Cr~{\sc i}, 
Cr~{\sc ii}, Mn~{\sc i}, Ni~{\sc i}, and Zn~{\sc i} were also
derived, with abundances of some species only obtained for stars with UVES
spectra, because of the larger spectral coverage. Corrections due to the
hyperfine structure (references in Gratton et al. 2003) were applied to Sc, V,
Mn, and Co. Abundances of Cu~{\sc i} were derived from spectrum synthesis, as
detailed in Carretta et al. (2011).

The concentration of the neutron-capture elements Y~{\sc ii}, Ba~{\sc ii}, 
La~{\sc ii}, Nd~{\sc ii} was derived mostly for stars with UVES spectra and
mostly from measurements of $EW$s. Results for Y and Ba were checked with
synthetic spectra using line lists from D'Orazi et al. (2013) and D'Orazi et al.
(2012), respectively. Abundances derived with the two methods are in very good
agreement.
Lanthanum abundances were obtained from $EW$s of 3-4 lines with transition parameters
from Sneden et al. (2003; see also Carretta et al. 2011).

The available Ba lines are all very strong and quite sensitive to the velocity
fields in the stellar atmospheres. As a consequence, a clear trend of Ba
abundances as a function of the microturbulence results when using the values of
$v_t$ derived using the weaker Fe lines, formed typically in deeper atmospheric
layers (see e.g. Carretta et al. 2013b for a discussion of a similar effect in
the analysis of NGC~362). To alleviate this problem, good results are
obtained by adopting the values of $v_t$ from the relation as a function of the
surface gravity provided by Worley et al. (2013) for giants in the metal-poor GC
M~15. When analyzed using these values and a constant metallicity equal to the
metal abundance derived for NGC~4833 (see next section), no trend is apparent
and we can safely explore the Ba abundances looking for intrinsic dispersion or
correlations (if any) with other elements.
The relation from Worley et al. was chosen since it was shown 
to efficiently remove any trend between Ba abundances and $v_t$ for bright stars
in a GC with metallicity comparable to that in NGC~4833. On average, the values
of $v_t$ from this relation are 0.26 kms$^{-1}$ higher than the values derived 
as described in the previous section for individual stars (with $rms=0.29$
kms$^{-1}$, 78 stars). However, for all other species the last method works
very well and we adopted for all elements except Ba this approach that
guarantees homogeneity with the more than 20 GCs analyzed by our group in this
FLAMES  survey.

\subsection{Metal abundances}

The mean metallicity we found for NGC~4833 from stars with high resolution 
UVES spectra is 
[Fe/H]$=-2.015 \pm0.004 \pm0.084$ dex ($rms=0.014$ dex, 12 stars) from neutral
species, where the
first error is from statistics and the second one refers to the systematic
effects, as estimated in the next section. From the large sample of stars with 
GIRAFFE spectra we derived a value of [Fe/H]$=-2.040 \pm0.003 \pm0.073$ dex 
($rms=0.024$ dex, 73 stars).

\begin{figure}
\centering 
\includegraphics[bb=78 171 460 691, clip,scale=0.52]{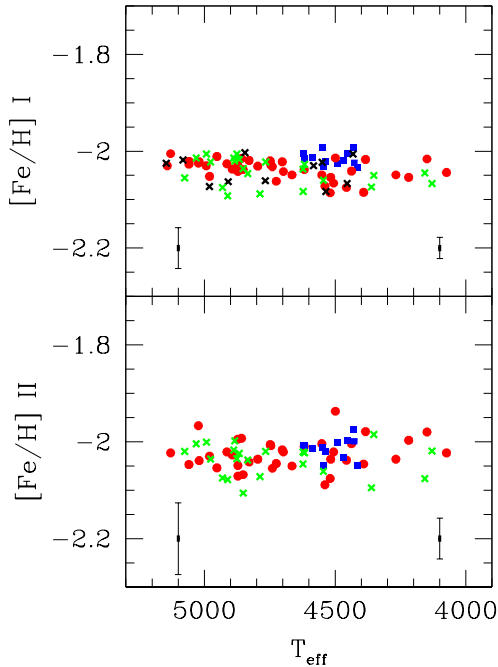}
\caption{Abundance ratios [Fe/H] {\sc i} (upper panel) and [Fe/H] {\sc ii}
(lower panel) as a function of T$_{\rm eff}$ for all analysed stars. Blue
squares are stars with UVES spectra, filled circles are those with GIRAFFE
spectra observed with both the HR11 and HR13 setups, and crosses indicate
stars observed with only the HR11 (black) or the HR13 (green) setup. Error bars
on the right and on the left side are star-to-star errors for targets observed
with UVES and GIRAFFE, respectively.}
\label{f:feteff48}
\end{figure}

The abundances of iron obtained from the singly ionized species are in excellent 
agreement with those from neutral lines: [Fe/H]{\sc ii}$=-2.014$  ($rms=0.021$
dex, 12 stars) from UVES and [Fe/H]{\sc ii}$=-2.030$ ($rms=0.033$ dex, 59 stars)
from GIRAFFE. The derived Fe abundances do not present any trend as a function
of the effective temperature, as shown in Fig.~\ref{f:feteff48}.

Our average metal abundance seems to be about 0.2-0.3 dex lower than most of
previous estimates in literature. From a preliminary analysis of RR Lyrae
variables, Murray and Darragh (2013) found [Fe/H]$=-1.67 \pm 0.13$ dex using
Fourier decomposition of the light curves.  We note, however, that often the
metallicities based on this method are higher than those derived from high
resolution spectroscopy, in particular in the low metallicity regime. For
example, [Fe/H]$=-1.98$ dex from RR Lyrae in M~15 (Garcia Lugo et al. 2007),
compared to [Fe/H]$=-2.32$ dex from high resolution spectroscopy (Carretta et
al. 2009c); [Fe/H]$=-1.23$ dex for M~5 from RR Lyrae (Kaluzny et al. 2000)
compared to $-1.34$ (Carretta et al. 2009c); [Fe/H]$=-2.11$ dex from variables in
M~30 (Kains et al. 2013), and $-2.34$ from Carretta et al. (2009c). 
The discrepancy is not limited to our group/analysis: Figuera Jaimes et al.
(2013) derived [Fe/H]$=-1.64$ from RR Lyraes in NGC~7492, while Cohen and
Melendez (2005) found $-1.82$ dex for this cluster from high resolution spectra.
Using high resolution spectra and MARCS models, Kraft and Ivans (2003) found
[Fe/H]$=-2.33$ dex and [Fe/H]$=-2.39$ dex for the metal-poor GCs M~30 and M~15,
respectively.

From CCD photometry Melbourne et al. (2000) derived a mean metallicity 
[Fe/H]$=-1.83 \pm0.14$ dex (and a reddening $E(B-V)=0.32 \pm0.03$). Early
detections based on high resolution spectra also obtained similar high values:
[Fe/H]$=-1.71$ dex from one star (Minniti et al. 1996) and [Fe/H]$=-1.74$ dex,
on average, from two stars (Gratton and Ortolani 1989). Moreover, Kraft
and Ivans (2003) found for this GC a metallicity [Fe/H]$=-2.06, -2.00, -2.04$
dex from MARCS, and Kurucz models with and without the overshooting option,
respectively.

\subsection{Error budget}

Our procedure for error estimates is amply described in Carretta et al. (2007b,
2009a,b) and will not be repeated here.
In the following we only provide the main tables of sensitivities of abundance
ratios to the adopted errors in the atmospheric parameters and $EW$s and the
final estimates of internal and systematic errors for all species analysed from
UVES and GIRAFFE spectra of stars in NGC~4833.

The sensitivities of derived abundances on the adopted atmospheric parameters 
were obtained by repeating our abundance analysis by changing only one 
atmospheric parameter each time for all stars in NGC~4833 (separately for the
UVES and the GIRAFFE samples). The sensitivity in each parameter was adopted as
the one corresponding to the average of all the sample.

The amount of the change in the input parameters T$_{\rm eff}$, $\log g$,
[A/H], and $v_t$ to compute the sensitivity of abundances to  
variations in the atmospheric parameters is shown in the
first line of the headers in Table~\ref{t:sensitivityu48}
and Table~\ref{t:sensitivitym48}, whereas the resulting 
response in abundance changes of all elements (the sensitivities) are shown 
in columns from 3 to 6 of these tables.

\setcounter{table}{2}
\begin{table*}
\centering
\caption[]{Sensitivities of abundance ratios to variations in the atmospheric
parameters, and to errors in the equivalent widths, and errors in abundances for
stars of NGC~4833 observed with UVES.}
\begin{tabular}{lrrrrrrrr}
\hline
Element     & Average  & T$_{\rm eff}$ & $\log g$ & [A/H]   & $v_t$    & EWs     & Total   & Total      \\
            & n. lines &      (K)      &  (dex)   & (dex)   &kms$^{-1}$& (dex)   &Internal & Systematic \\
\hline        
Variation&             &  50           &   0.20   &  0.10   &  0.10    &         &         &            \\
Internal &             &   4           &   0.04   &  0.01   &  0.10    & 0.01    &         &            \\
Systematic&            &  57           &   0.06   &  0.08   &  0.03    &         &         &            \\
\hline
$[$Fe/H$]${\sc  i}& 39 &    +0.073     & $-$0.016 &$-$0.015 & $-$0.016 & 0.014  &0.022    &0.084	\\
$[$Fe/H$]${\sc ii}&  5 &  $-$0.017     &   +0.079 &  +0.018 & $-$0.005 & 0.038  &0.042    &0.032	\\
$[$O/Fe$]${\sc  i}&  1 &  $-$0.050     &   +0.091 &  +0.039 &   +0.014 & 0.085  &0.088    &0.090	\\
$[$Na/Fe$]${\sc i}&  2 &  $-$0.038     & $-$0.030 &$-$0.003 &   +0.011 & 0.060  &0.061    &0.096	\\
$[$Mg/Fe$]${\sc i}&  1 &  $-$0.029     &   +0.000 &  +0.002 &   +0.011 & 0.085  &0.086    &0.072	\\
$[$Al/Fe$]${\sc i}&  1 &  $-$0.034     &   +0.002 &  +0.003 &   +0.013 & 0.085  &0.086    &0.105	\\
$[$Si/Fe$]${\sc i}&  4 &  $-$0.050     &   +0.025 &  +0.011 &   +0.013 & 0.043  &0.045    &0.059	\\
$[$Ca/Fe$]${\sc i}& 15 &  $-$0.016     & $-$0.006 &$-$0.003 &   +0.003 & 0.022  &0.022    &0.018	\\
$[$Sc/Fe$]${\sc ii}& 6 &    +0.011     & $-$0.007 &  +0.003 &   +0.011 & 0.035  &0.035    &0.029	\\
$[$Ti/Fe$]${\sc i}&  5 &    +0.019     & $-$0.005 &$-$0.002 & $-$0.014 & 0.038  &0.040    &0.014	\\
$[$Ti/Fe$]${\sc ii}& 9 &    +0.025     & $-$0.013 &$-$0.002 & $-$0.002 & 0.028  &0.032    &0.023	\\
$[$V/Fe$]${\sc i} &  2 &    +0.016     & $-$0.006 &$-$0.002 &   +0.013 & 0.060  &0.062    &0.019	\\
$[$Cr/Fe$]${\sc i}&  9 &    +0.013     & $-$0.011 &$-$0.007 & $-$0.009 & 0.028  &0.030    &0.017	\\
$[$Cr/Fe$]${\sc ii}& 4 &  $-$0.001     & $-$0.011 &$-$0.009 &   +0.000 & 0.043  &0.043    &0.013	\\
$[$Mn/Fe$]${\sc i}&  2 &  $-$0.012     & $-$0.000 &  +0.001 &   +0.014 & 0.060  &0.062    &0.015	\\
$[$Co/Fe$]${\sc i}&  1 &  $-$0.014     &   +0.001 &  +0.003 &   +0.013 & 0.085  &0.086    &0.020	\\
$[$Ni/Fe$]${\sc i}&  8 &    +0.005     &   +0.008 &  +0.003 &   +0.007 & 0.030  &0.031    &0.007	\\
$[$Cu/Fe$]${\sc i}&  1 &    +0.010     &   +0.003 &  +0.001 &   +0.010 & 0.085  &0.086    &0.021	\\
$[$Zn/Fe$]${\sc i}&  1 &  $-$0.075     &   +0.056 &  +0.022 & $-$0.001 & 0.085  &0.085    &0.088	\\
$[$Y/Fe$]${\sc ii}&  9 &    +0.031     & $-$0.012 &  +0.000 & $-$0.013 & 0.028  &0.031    &0.040	\\
$[$Ba/Fe$]${\sc ii}& 3 &    +0.043     & $-$0.010 &  +0.002 & $-$0.075 & 0.049  &0.090    &0.065        \\
$[$La/Fe$]${\sc ii}& 3 &    +0.040     & $-$0.009 &  +0.003 &   +0.001 & 0.049  &0.049    &0.046        \\
$[$Nd/Fe$]${\sc ii}& 4 &    +0.041     & $-$0.009 &  +0.003 & $-$0.000 & 0.043  &0.043    &0.048        \\
\hline
\end{tabular}
\label{t:sensitivityu48}
\end{table*}

\begin{table*}
\centering
\caption[]{Sensitivities of abundance ratios to variations in the atmospheric
parameters and to errors in the equivalent widths, and errors in abundances for
stars of NGC~4833 observed with GIRAFFE}
\begin{tabular}{lrrrrrrrr}
\hline
Element     & Average  & T$_{\rm eff}$ & $\log g$ & [A/H]   & $v_t$    & EWs    & Total   & Total      \\
            & n. lines &      (K)      &  (dex)   & (dex)   &kms$^{-1}$& (dex)  &Internal & Systematic \\
\hline        
Variation&             &  50           &   0.20   &  0.10   &  0.10    &        &	  &	       \\
Internal &             &   4           &   0.04   &  0.02   &  0.22    & 0.02   &	  &	       \\
Systematic&            &  57           &   0.06   &  0.07   &  0.03    &        &	  &	       \\
\hline
$[$Fe/H$]${\sc  i}& 20 &    +0.064     & $-$0.011 &$-$0.011 & $-$0.016 & 0.023  &0.042    &0.073     \\
$[$Fe/H$]${\sc ii}&  2 &  $-$0.016     &   +0.077 &  +0.011 & $-$0.004 & 0.071  &0.074    &0.030     \\
$[$O/Fe$]${\sc  i}&  1 &  $-$0.040     &   +0.085 &  +0.031 &   +0.018 & 0.101  &0.110    &0.066     \\
$[$Na/Fe$]${\sc i}&  2 &  $-$0.034     & $-$0.026 &  +0.004 &   +0.013 & 0.071  &0.077    &0.055     \\
$[$Mg/Fe$]${\sc i}&  1 &  $-$0.027     &   +0.001 &  +0.002 &   +0.013 & 0.101  &0.105    &0.039     \\
$[$Si/Fe$]${\sc i}&  3 &  $-$0.046     &   +0.021 &  +0.009 &   +0.015 & 0.058  &0.067    &0.053     \\
$[$Ca/Fe$]${\sc i}&  4 &  $-$0.015     & $-$0.006 &$-$0.001 &   +0.001 & 0.051  &0.051    &0.017     \\
$[$Sc/Fe$]${\sc ii}& 4 &  $-$0.052     &   +0.082 &  +0.026 &   +0.011 & 0.051  &0.059    &0.064     \\
$[$Ti/Fe$]${\sc i}&  3 &    +0.004     & $-$0.004 &  +0.001 &   +0.015 & 0.058  &0.067    &0.007     \\
$[$V/Fe$]${\sc i} &  3 &    +0.021     & $-$0.008 &$-$0.003 &   +0.020 & 0.058  &0.073    &0.025     \\
$[$Cr/Fe$]${\sc i}&  1 &    +0.011     & $-$0.006 &$-$0.001 &   +0.021 & 0.101  &0.111    &0.018     \\
$[$Co/Fe$]${\sc i}&  1 &  $-$0.001     &   +0.003 &  +0.005 &   +0.023 & 0.101  &0.113    &0.015     \\
$[$Ni/Fe$]${\sc i}&  3 &  $-$0.001     &   +0.008 &  +0.004 &   +0.013 & 0.058  &0.065    &0.006     \\
$[$Ba/Fe$]${\sc ii}& 1 &  $-$0.033     &   +0.077 &  +0.024 & $-$0.067 & 0.101  &0.179    &0.056     \\
\hline
\end{tabular}
\label{t:sensitivitym48}
\end{table*}

The averages of all measured elements with their $r.m.s.$ scatter are listed 
in  Table~\ref{t:meanabu48}. Derived atmospheric parameters and Fe abundances for
individual stars in NGC~4833 are in Table~\ref{t:atmpar48}; abundances of
proton-capture, $\alpha$-capture, Fe-peak and neutron-capture elements are
provided in Tables~\ref{t:proton48}, \ref{t:alpha48}, \ref{t:fegroup48}, 
\ref{t:neutron48} and \ref{t:ba48}, respectively. These tables are only available 
in electronic form: a few lines are given for guidance. Upon request by the
referee, we also report in the last two columns of Table~\ref{t:proton48} the
average and rms scatter of the [Na/Fe] ratio in LTE for individual stars, to
give an idea of the adopted NLTE corrections.

\begin{table}
\centering
\caption{Mean abundances from UVES and GIRAFFE }
\begin{tabular}{lcc}
\hline
                     &               &               \\
Element              & UVES	     & GIRAFFE       \\
                     &n~~   avg~~  $rms$ &n~~	avg~~  $rms$ \\        
\hline
$[$O/Fe$]${\sc i}    &12   +0.17 0.22 &51   +0.25 0.29  \\
$[$Na/Fe$]${\sc i}   &12   +0.52 0.29 &52   +0.46 0.27  \\
$[$Mg/Fe$]${\sc i}   &12   +0.27 0.22 &44   +0.36 0.15  \\
$[$Al/Fe$]${\sc i}   &12   +0.90 0.34 &                 \\
$[$Si/Fe$]${\sc i}   &12   +0.47 0.04 &66   +0.46 0.05  \\
$[$Ca/Fe$]${\sc i}   &12   +0.35 0.01 &73   +0.35 0.02  \\
$[$Sc/Fe$]${\sc ii}  &12 $-$0.04 0.01 &73 $-$0.04 0.02  \\
$[$Ti/Fe$]${\sc i}   &12   +0.18 0.02 &53   +0.17 0.02  \\
$[$Ti/Fe$]${\sc ii}  &12   +0.23 0.01 &                 \\
$[$V/Fe$]${\sc i}    &12 $-$0.08 0.01 &24 $-$0.10 0.02  \\
$[$Cr/Fe$]${\sc i}   &12 $-$0.24 0.02 &24 $-$0.20 0.04  \\
$[$Cr/Fe$]${\sc ii}  &12   +0.01 0.05 &  \\
$[$Mn/Fe$]${\sc i}   &12 $-$0.54 0.01 &  \\
$[$Fe/H$]${\sc i}    &12 $-$2.02 0.01 &73 $-$2.04 0.02  \\
$[$Fe/H$]${\sc ii}   &12 $-$2.01 0.02 &59 $-$2.03 0.03  \\
$[$Co/Fe$]${\sc i}   & ~8 $-$0.03 0.03 & ~7 $-$0.07 0.04  \\
$[$Ni/Fe$]${\sc i}   &12 $-$0.18 0.01 &68 $-$0.18 0.03  \\
$[$Cu/Fe$]${\sc i}   &12 $-$0.80 0.09 &                 \\ 
$[$Zn/Fe$]${\sc i}   &12   +0.07 0.03 &                 \\  
$[$Y/Fe$]${\sc ii}   &12 $-$0.15 0.06 &                 \\ 
$[$Ba/Fe$]${\sc ii}  &12 $-$0.06 0.07 &62 $-$0.20 0.14  \\ 
$[$La/Fe$]${\sc ii}  &12   +0.05 0.03 &      \\ 
$[$Nd/Fe$]${\sc ii}  &12   +0.42 0.04 &      \\ 
\hline
\end{tabular}
\label{t:meanabu48}
\end{table}

\section{Results}

\subsection{The Na-O anticorrelation in NGC~4833}

\begin{figure}
\centering 
\includegraphics[scale=0.42]{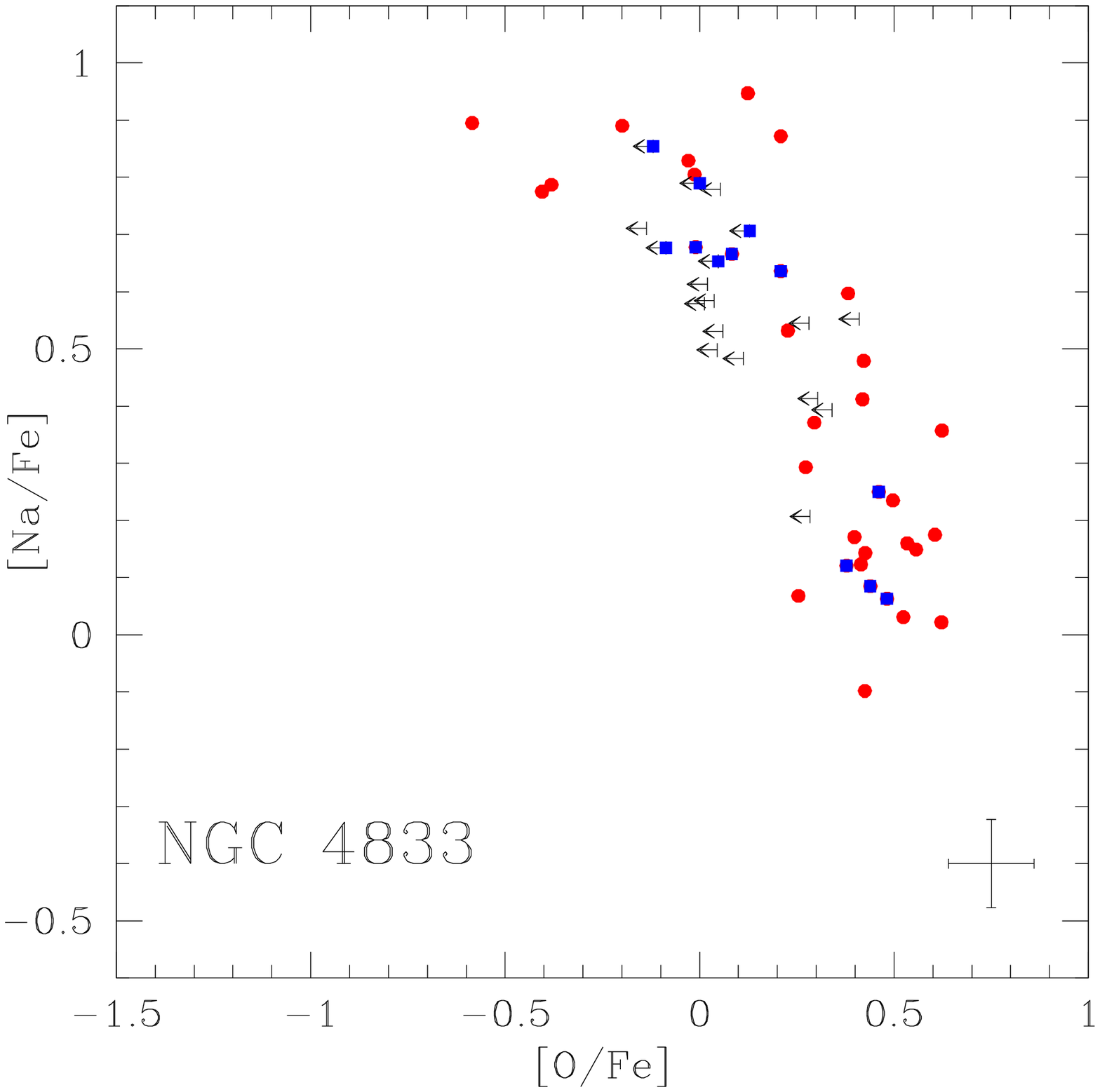}
\caption{The Na-O anti-correlation observed in NGC~4833. Blue squares are stars
observed with UVES, while red circles indicate stars with GIRAFFE spectra.
Upper limits in O are shown as arrows, and star-to-star (internal) error bars
are plotted.}
\label{f:m48antiu}
\end{figure}

After combining the UVES and GIRAFFE datasets and taking into account stars
observed with both instruments, we ended with 61 stars with O abundances (40
actual detections and 21 upper limits) and 60 stars with Na abundances.  The
Na-O anticorrelation in NGC~4833 rests on 51 giants with both O and Na, and is
shown in Fig.~\ref{f:m48antiu}, with star-to-star errors relative to the GIRAFFE
dataset. Internal error bars for the UVES sample are slightly smaller (see
Table~\ref{t:sensitivityu48}).

It is not easy to judge whether stars in NGC~4833 are grouped into discrete
populations with homogeneous composition, along the Na-O anticorrelation, mostly
formed with the stars observed with GIRAFFE. Similar large samples are better
suited to quantify the extension of this feature, but the associated internal
errors may smear possible groups, although some sub-divisions are recognizable
in Fig.~\ref{f:m48antiu}. The limited sample of giants observed
with UVES seems to be better suited to this task.

\begin{figure}
\centering 
\includegraphics[bb=44 187 450 701, clip, scale=0.55]{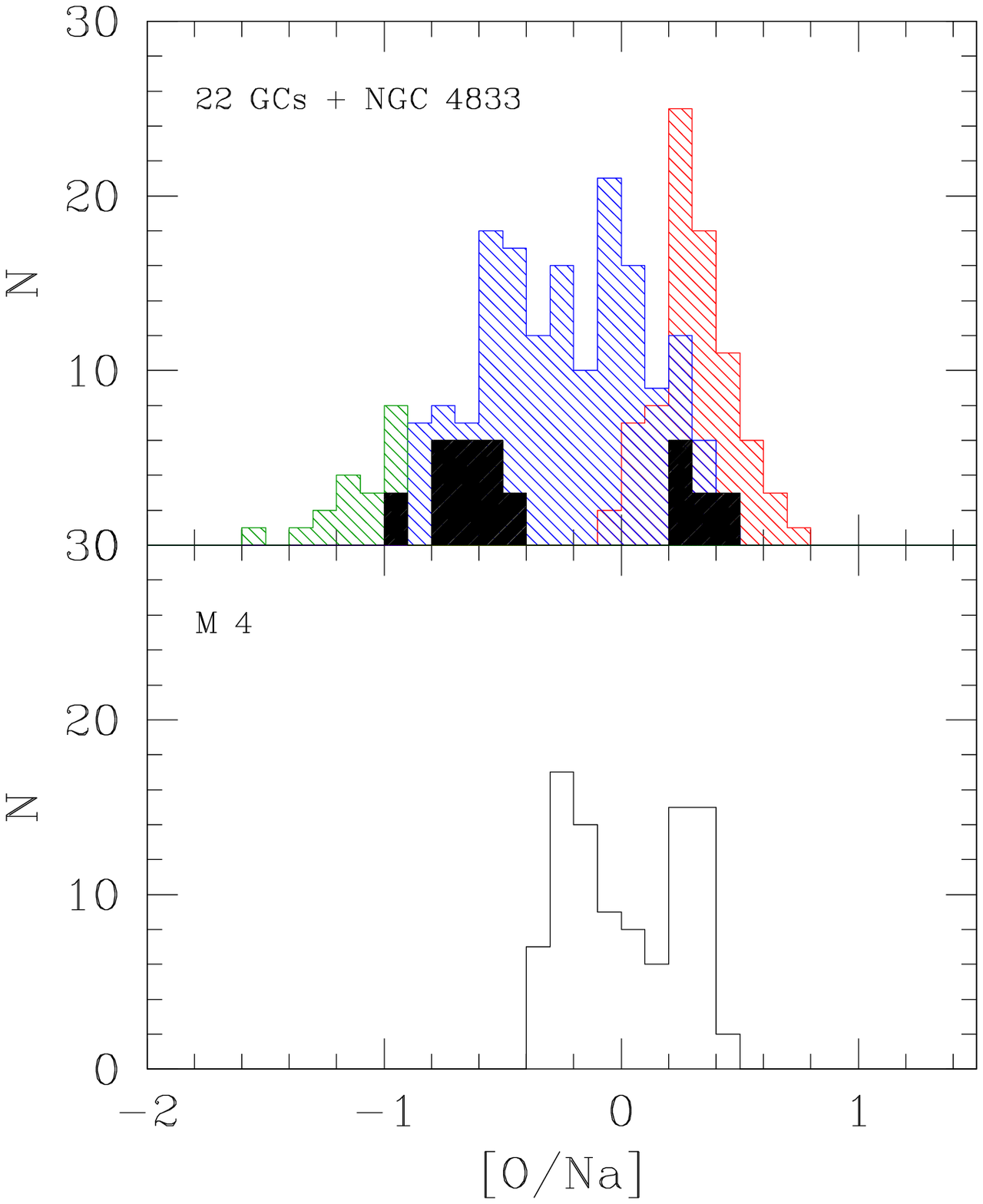}
\caption{Distribution of the [O/Na] abundance ratios in several GCs from stars
observed with the high resolution UVES spectrograph. In the upper panel, about
250 RGB stars in 23 GCs from our FLAMES survey are plotted. Colour coding is
red, blue, and green for the P, I, and E components defined in Carretta et al.
(2009a). In black we plotted the distribution for NGC~4833 from the present
work, with the numbers of stars multiplied by 3 to improve
the visibility. Lower panel: distribution of [O/Na] from UVES spectra of about
100 giants in M~4 from Marino et al. (2008).}
\label{f:histo2}
\end{figure}

The 12 stars with UVES spectra are clearly clustered in two clearly separated
groups: one with abundances typical of the pattern established by core-collapse
supernovae nucleosynthesis (high O and low Na), also shared by field stars of
similar metallicity (e.g. Gratton et al. 2000), the other with low-O/high-Na
abundances, whose counterparts are rarely observed in Galactic field stars.
To check that this occurrence in NGC~4833 is not a spurious effect due to the
limited number of giants in the UVES sample, we plotted in Fig.~\ref{f:histo2}
(upper panel) the distribution of the [O/Na] ratios from our total sample of
about 250 RGB stars observed with UVES in 22 GCs of the Milky Way in our FLAMES
survey. The colour coding corresponds to the division of stars into the
primordial (P) component of first generation stars, and to the two fractions of
stars with intermediate (I) and extreme (E) composition within the second
stellar generation in GCs, as defined in Carretta et al. (2009a) from their
location along the Na-O anticorrelation. Our UVES sample in NGC~4833 (shown in
Fig.~\ref{f:histo2} with number counts multiplied by a factor 3 to improve
clarity) clearly splits into two groups, roughly coincident with the first and
second generation stars. 
As a comparison, in the lower panel we also plot the only other large sample in
an individual GC based on high resolution spectra, the about 100 RGB stars
observed in M~4 by Marino et al. (2008). An offset of 0.1 dex was arbitrarily 
subtracted to their [O/Na] values to bring them on our abundance scale. In M~4,
with its short anticorrelation, the extreme component of second generation stars
is obviously not present.

Using the quantitative criteria introduced by Carretta et al. (2009a) we can 
use O and Na abundances to quantify the fraction of the different stellar
generations. From the total sample of 51 stars with O and Na we found that the 
fractions of P, I, and E stars for NGC~4833 are
$31\pm8\%$, $59\pm11\%$, and $10\pm4\%$,  respectively. The fraction of first
generation stars is similar to the one (about one third) typical of the
overwhelming majority of Galactic GCs. On the other hand, the  fraction of
second generation E stars with extremely modified composition is quite large in
NGC~4833. As a comparison, the E fraction in GCs like NGC~4590 (M~68), NGC~6809
(M~55), NGC~7078 (M~15), NGC~7099 (M~30), bracketing NGC~4833 in mass and
metallicity, does not exceed 2-3\% (being formally absent in M~15 and M~68,
Carretta et al. 2009a,b).
Among the observed RGB stars in NGC~4833 there is apparently no statistically
significant segregation in radial distance from the cluster centre for the P, I,
and E components. Our sample of member stars is however all confined within 
two half-mass radii and may be not the optimal sample for this kind of analysis.
Large photometric databases are better suited to study possibile difference of
radial concentration of different stellar generations.

\begin{figure}
\centering 
\includegraphics[bb=19 145 582 470, clip, scale=0.42]{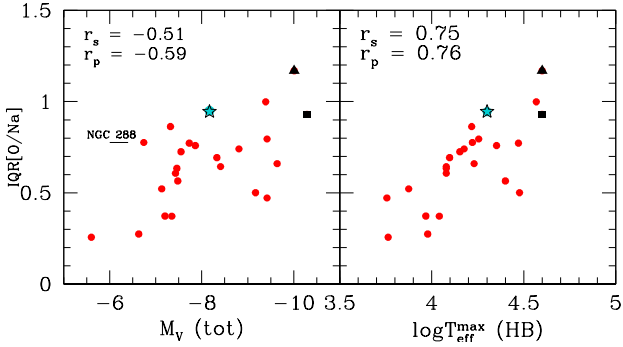}
\caption{IQR[O/Na] ratios for NGC~4833 (star symbol) and other
Galactic GCs as a function of the total cluster absolute magnitude $M_V$ 
(left panel). The other clusters are $\omega$ Cen (filled square, Johnson and 
Pilachowski 2010), M~54 (filled triangle, Carretta et al. 2010b), and other GCs
from our FLAMES survey (filled circles, Carretta et al. 2009a, 2011, 2013b). In
the right panel the IQR[O/Na] is shown as a function of the maximum temperature
on the horizontal branch, from Recio-Blanco et al. (2006).
In each panel, the Spearman rank correlation coefficient
($r_s$) and the Pearson's correlation coefficient ($r_p$) are reported.} 
\label{f:iqrmvteff48}
\end{figure}

The interquartile range (IQR) for the ratio [O/Na] is a very useful measurement
to quantify the extension of the Na-O anticorrelation (Carretta 2006). From our
large sample we found that IQR[O/Na]=0.945 dex in NGC~4833.
Therefore, this cluster joins the ensemble of other GCs producing a nice
correlation with the cluster total mass (represented by the proxy of the total 
absolute magnitude, $M_V=-8.16$ for NGC~4833 from Harris 1996), established in 
Carretta et al. (2010a) and reproduced in Fig.~\ref{f:iqrmvteff48}, left panel.
The location of NGC~4833 in this plot seems to be on the upper envelope of the
relation defined by the bulk of the other GCs. The position of another cluster
sharing a similar position (NGC~288) is also indicated. This occurrence will be
discussed in Section 5.

Recio-Blanco et al. (2006) computed the maximum temperature reached along the
horizontal branch (HB) in NGC~4833: $\log T_{\rm eff}=4.301$. The correlation
between this parameter and the extension of the Na-O anticorrelation, discovered
by Carretta et al. (2007c), is updated and shown in the right panel of
Fig.~\ref{f:iqrmvteff48}. Once again, NGC~4833 seems to lie at the upper
envelope of the relation.

\subsection{Other proton-capture elements}

Apart from the case of O and Na, significant star-to-star abundance variations
are detected for other proton-capture elements in giants in NGC~4833. In
particular, we found that the [Mg/Fe] abundance ratio shows an unusually large 
spread in this cluster, with peak-to-peak variations of more than 0.5 dex.

The reality of the intrinsic scatter in Mg is immediately evident comparing the
estimated internal error for the UVES sample (0.086 dex) to the observed $rms$
scatter in [Mg/Fe] (0.223 dex): the cosmic spread in Mg among RGB stars in
NGC~4833 is significant at almost 3$\sigma$ level.

\begin{figure}
\centering 
\includegraphics[scale=0.45]{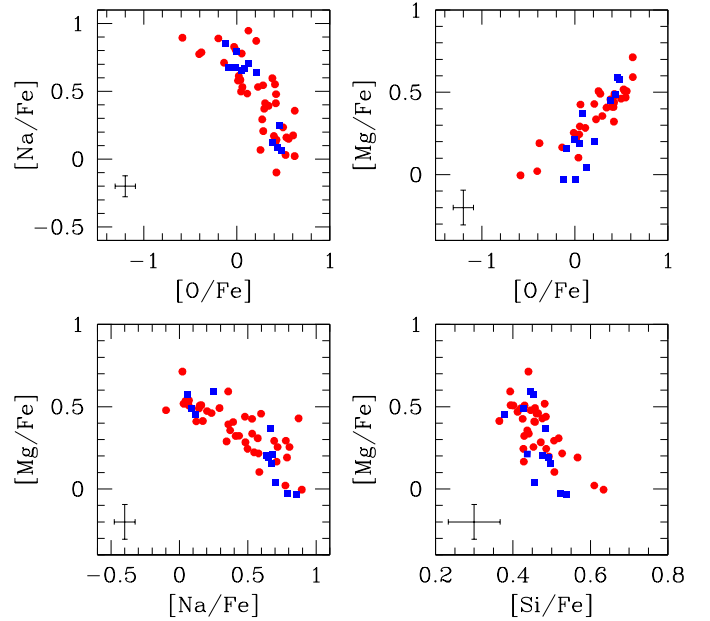}
\caption{Relations among proton-capture elements O, Na, Mg, and Si in 
NGC~4833. Red circles refer to stars with GIRAFFE spectra, blue squares
indicate stars observed with UVES. The internal errorbars plotted in each panel
are those relative to the GIRAFFE sample.}
\label{f:light1u48}
\end{figure}

The run of [Mg/Fe] ratios as a function of the abundance of the proton-capture
elements O, Na, and Si is shown in Fig.~\ref{f:light1u48}, together with the
classical Na-O anticorrelation. The error bars indicate star-to-star errors and
are referred to the GIRAFFE sample; internal errors for star of the UVES sample
are usually smaller (see Table~\ref{t:sensitivityu48}). The Mg abundance is
correlated to that of O and anticorrelated with species enhanced in the network
of proton-capture reactions, namely Na and Si. We retrieved this pattern from
both the datasets observed with UVES and with GIRAFFE.

\begin{figure}
\centering 
\includegraphics[bb=44 187 450 701, clip, scale=0.55]{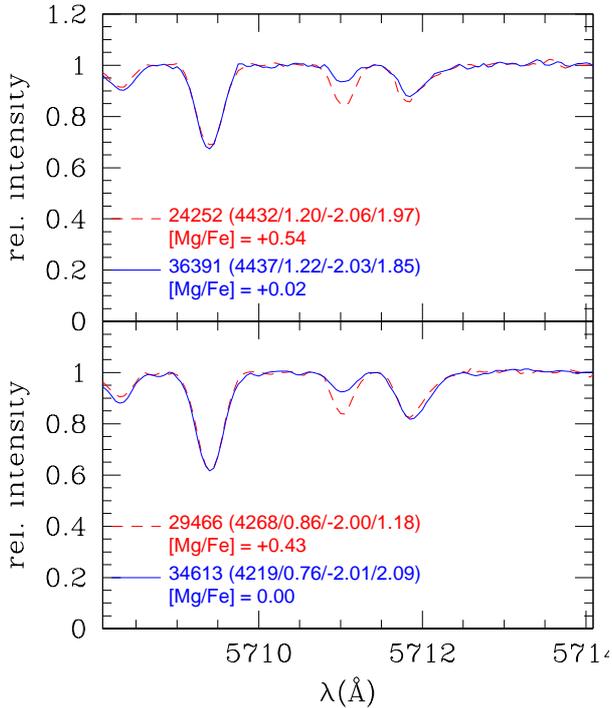}
\caption{Comparison of two pairs of Mg-poor and Mg-normal stars with similar 
atmospheric parameters in the Mg~{\sc i} 5711.09~\AA\ spectral region.
Atmospheric parameters $T_{\rm eff}$, $\log g$, metallicity, $v_t$, and the
[Mg/Fe] abundance ratios of stars are indicated in each panel.}
\label{f:m48mglow}
\end{figure}

In the case of the GIRAFFE sample, the observed spread in Mg (0.151
dex) does not formally exceed too much the associated internal error (0.105 dex), likely
due to the lower resolution of the spectra and the extension of the GIRAFFE
sample to warmer giants, with weaker lines.
However, even in this case there is no doubt that Mg variations in NGC~4833 are
real. In Fig.~\ref{f:m48mglow} the HR11 spectra of the two stars with the
lowest Mg abundances in the GIRAFFE sample (star 36391, with [Mg/Fe]=+0.02 dex,
and star 34613 with [Mg/Fe]=0.00) are compared to the spectra of two other stars
with similar atmospheric parameters, but quite different (much higher) Mg
abundances.

This direct comparison, free from any uncertainties related to the abundance
analysis, robustly corroborates our findings: there is a large spread in Mg in
NGC~4833, where giants with a solar [Mg/Fe] ratio stand side to side with
stars having a normal [Mg/Fe] ratio appropriate for metal-poor halo stars.

Large depletions of Mg due to the action of proton-capture reactions in
H-burning at high temperature should have two main consequences: produce Al
through the Mg-Al cycle and, were the burning temperatures high enough, slightly 
enhance the abundance of Si through the mechanism of the leakage from the
Mg-Al cycle on $^{28}$Si (Karakas and Lattanzio 2003).

\begin{figure}
\centering 
\includegraphics[scale=0.45]{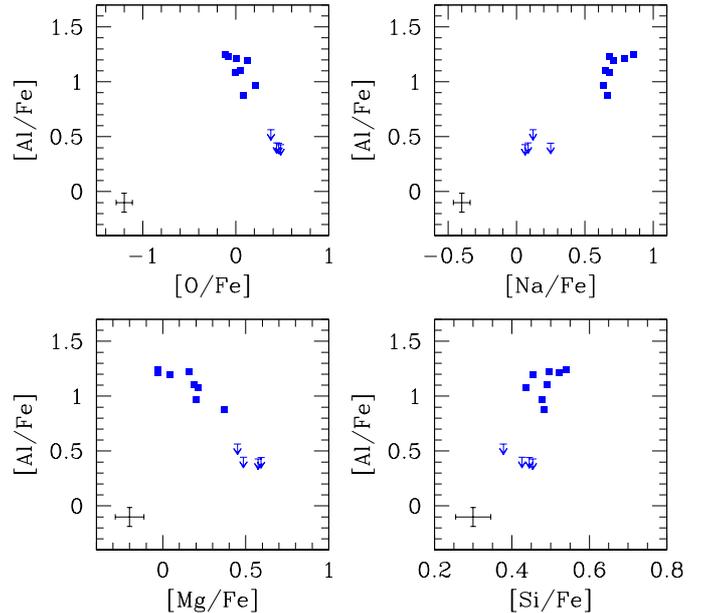}
\caption{Relations of Al with the proton-capture elements O, Na, Mg, and Si in
giants of NGC~4833 observed with UVES spectra. Error bars represent internal
errors associated to the UVES sample. The upper limits in  Al abundances 
are indicated by arrows.}
\label{f:light2u48}
\end{figure}

Abundances of Al can be obtained from the UVES spectra; in four stars out 
of 12 observed with UVES only upper limits could be derived.
Nevertheless, the derived values do not
show any trend as a function of the effective temperature and 
their position reveals instead clear
patterns (Fig.~\ref{f:light2u48}), with the typical correlations and
anticorrelations among elements produced or destroyed, respectively, by the
interplay of the Ne-Na and Mg-Al cycles (Denisenkov and Denisenkova 1989, Langer
et al. 1993). Moreover, the sample, albeit limited, splits into two clearly
separated groups.

\begin{figure}
\centering 
\includegraphics[bb=19 406 576 700, clip, scale=0.42]{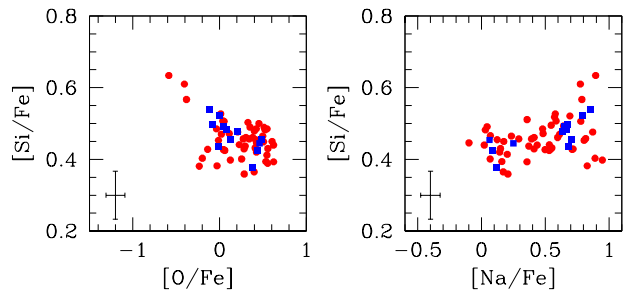}
\caption{Relations among proton-capture element ratios [Si/Fe], [O/Fe], and
[Na/Fe] in NGC~4833. Blue squares indicate stars of the UVES sample, red
circles are for stars observed with GIRAFFE. The internal error bars refer to
the latter sample.}
\label{f:light3u48}
\end{figure}

The Mg-Si anticorrelation in Fig.~\ref{f:light1u48}, together with the Si-Al
correlation (Fig.~\ref{f:light2u48}), already show that some Si production
occurred in the polluters of the first stellar generation in NGC~4833. To
further support this evidence, we plot in  Fig.~\ref{f:light3u48} the [Si/Fe]
ratios as a function of the O and Na values.
The evidence of a Si-O anticorrelation is not robust for the GIRAFFE sample, but
is clear for the more limited UVES sample. The correlation between Si and Na is
well represented in both samples. The two stars with the lowest Mg abundances
(Fig.~\ref{f:light1u48} and Fig.~\ref{f:m48mglow}) are also those showing the 
highest Si abundances, leaving no doubt that trends in Si and Mg are due to a
common mechanism.

\subsection{Other elements}

\begin{figure}
\centering 
\includegraphics[scale=0.42]{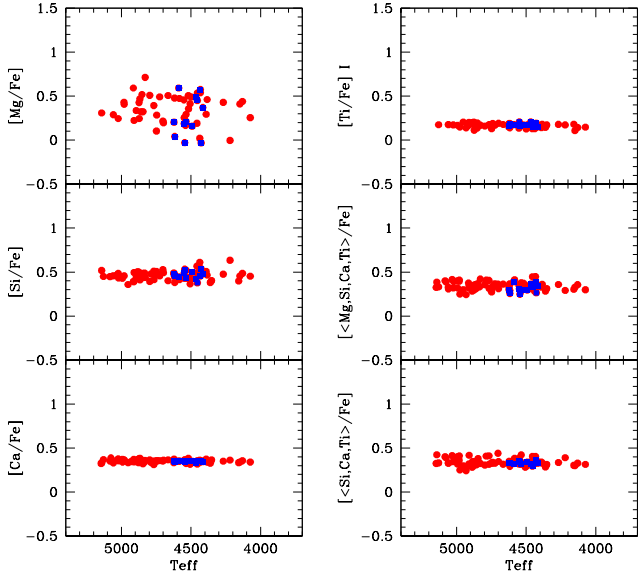}
\caption{Abundance ratios of $\alpha-$elements Mg, Si, Ca, Ti~{\sc i} as a
function of the effective temperature. The average of [$\alpha$/Fe] ratios are
shown  in the last two panels on the right column (including and excluding the
Mg abundance from the mean, respectively). Blue squares are UVES stars. Internal
error bars are provided in Tables~\ref{t:sensitivityu48} and
~\ref{t:sensitivitym48}.}
\label{f:alpmu48}
\end{figure}

The pattern of the $\alpha-$elements measured in NGC~4833 is shown as a function
of the effective temperatures for indivisual stars in Fig.~\ref{f:alpmu48},
including also species like Mg and Si involved in the proton-capture reactions
discussed in the previous section.

The vertical scale, bracketing the range of [Mg/Fe] ratios, is the same
for all the elements to effectively show the large intrinsic dispersion of Mg
and, partly, of Si with respect to other $\alpha-$elements with no intrinsic
scatter in NGC~4833.

\begin{figure}
\centering 
\includegraphics[scale=0.42]{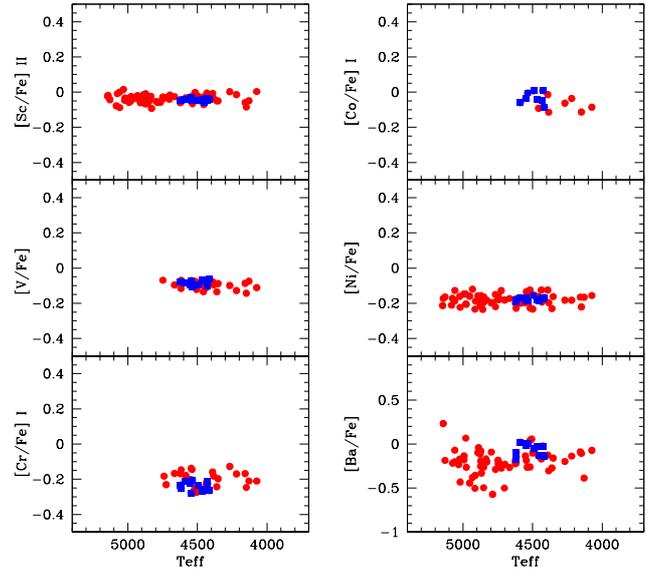}
\caption{Abundance ratios of elements of the Fe-peak (Sc~{\sc ii}, V, Cr, 
Co, Ni) and of the neutron-capture element Ba as a function of the effective 
temperature. Blue squares are UVES stars. Note the different vertical scale for
the panel with Ba abundances. Internal error bars are provided in 
Tables~\ref{t:sensitivityu48} and ~\ref{t:sensitivitym48}.} 
\label{f:fegmu48}
\end{figure}

The run of elements of the Fe-peak Sc, V, Cr, Co, and Ni as a function of the
temperature is shown in Fig.~\ref{f:fegmu48} for individual stars in NGC~4833
from the UVES and GIRAFFE samples. These elements present no surprise; they 
track iron, as in most GCs, with no trend as a function of the
$T_{\rm eff}$.
The elements Mn and Cu, only available for stars with UVES spectra, show the
underabundance typical of metal-poor GCs (e.g. Simmerer et al. 2003, Sobeck et
al. 2006).

In the right-bottom panel of Fig.~\ref{f:fegmu48} we plot the abundance ratios
of Ba, the only neutron-capture element available for a large sample of stars in
NGC~4833. As explained in Section~3.2, our finally adopted Ba abundances, displayed
in this panel, are those obtained by using for all stars a fixed model metal 
abundance ([Fe/H]$=-2.02$ dex, the average from the UVES spectra), and the
microturbulence from the relation $v_t=2.386-0.3067\log g$ derived by Worley et
al. (2013) for giants in the metal-poor GC M~15.

As recently shown in Worley et al. (2013) and Carretta et al. (2013a), this 
approach is quite effective in eliminating any trends of Ba abundances as a
function of $v_t$ and in reducing the ensueing spurious large scatters of the
average. Only for Ba we then adopt these values, our intent being simply to
state that this neutron-capture element from $s-$process in NGC~4833 (i) does
not have an intrinsic dispersion (compare the $rms$ scatters of the means in
Table~\ref{t:meanabu48} with the internal errors in Table~\ref{t:sensitivityu48}
and \ref{t:sensitivitym48}), and (ii) there is no relation between Ba and
elements involved in proton-capture reactions.

We cannot estimate from our data the relative contribution of the $r-$ and
$s-$process of neutron capture, because we did not measure a reliable abundance
for the typical species that, like Eu, primarily sample an almost pure
$r-$process nucleosynthesis at all metal abundances. 
For the [Ba/Y] ratio we found for NGC~4833 an average value of 0.09 dex
($rms=0.06$ dex, 12 stars) which is perfectly compatible with the ratios of
field stars and GCs of similar metallicity (see e.g. Venn et al. 2004).

As for Ba, we found no correlation or anti-correlation whatsoever between
the abundances of the $s-$process element La and the abundances of
proton-capture elements.

\section{Discussion}

\begin{figure}
\centering 
\includegraphics[scale=0.40]{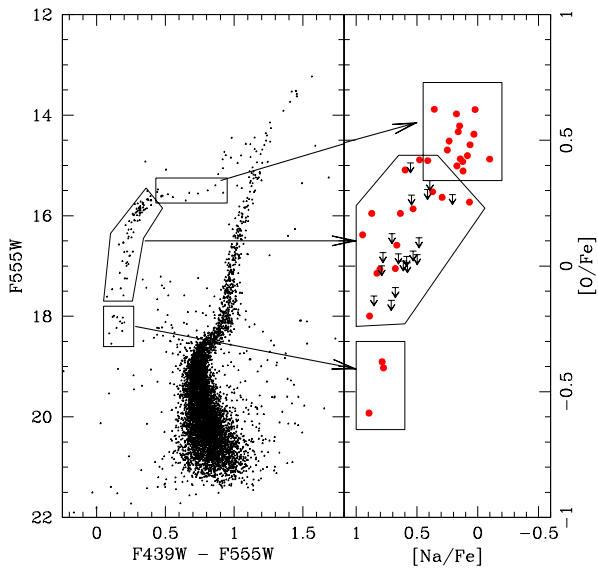}
\caption{$HST$ colour-magnitude CMD of NGC~4833 from the snapshot survey by
Piotto et al. (2002, left): the selected regions along the distribution of HB
stars are tentatively associated to different groups along the Na-O anticorrelation from the
present work (right panel).}
\label{f:hbvsnao}
\end{figure}

The Na-O anticorrelation among RGB stars in NGC~4833 reaches a quite long
extension. This result must not come unexpected. A statistically robust
relation between the extent of the Na-O anticorrelation and the hottest point
along the HB is well known (Carretta et al. 2007c, 2010a).
On the other hand, the HB in NGC~4833 presents a long blue tail, which stands out
clearly, free of field contamination, in particular when high resolution $HST$
imaging is used to construct the CMD (Fig.~\ref{f:hbvsnao}, left panel). 
Therefore, it would have been easy to predict a long Na-O anticorrelation in
this cluster, which is exactly what we found with the current analysis.

It is tempting to associate the groups selected on the HB in the left panel of 
Fig.~\ref{f:hbvsnao} to the RGB stars distributed in the Na-O anticorrelation
(right panel of the same figure). After all, RGB stars should end up on the HB
after igniting He burning at the centre. Qualitatively, the tentative
correspondences illustrated in Fig.~\ref{f:hbvsnao} may work well, provided that
no strong radial gradients are present between our spectroscopic sample and the
photometric sample, which cover different parts of the cluster. More
quantitative relations must of course await spectroscopic abundance analysis of
{\it in situ} HB stars.

However, the major peculiarity we uncovered in NGC~4833 is maybe the large
spread in Mg, whose abundance is clearly affected by large changes with respect
to the usual plateau established by supernovae nucleosynthesis. As discussed in
previous sections, we presented proofs that in this cluster also the Si 
abundance is partially modified by proton-capture reactions. This phenomenon was
first observed by Yong et al. (2005) in NGC~6752, another metal-deficient GC
with a long blue tail on the HB, and afterward individuated through Si-Al
correlations or Si-Mg anticorrelations in a number of other GCs (Carretta et al.
2009b). It is worth noting that in our large FLAMES survey we found significant
changes to the Mg (and Si) abundances only in massive and/or metal-poor GCs.

The leakage from the Mg-Al cycle on $^{28}$Si puts a strong constraint on the 
temperature at which H-burning occurred in the stars responsible for polluting
the intracluster gas, because the reaction producing
$^{28}$Si becomes dominant when T$_6 \sim 65$ K (Arnould et al. 1999, where the
temperature is expressed in millions of Kelvin).

Hence, although this constraint does not allow to discriminate the type of stars
providing the raw material for the formation of the second generation (see
Prantzos et al. 2007, their figure 8), a logical question is whether NGC~4833 is
another case like NGC~2419, although scaled down in amount of the involved
elemental variations. In the distant halo cluster NGC~2419, the third most
massive GC in our Galaxy, Cohen and Kirby (2012) and Mucciarelli et al. (2012)
discovered a double population of stars on the RGB. One group is made of 
Mg-normal giants with nearly solar abundance of potassium, whereas the other
includes stars with a huge depletion of Mg (and large enhancement of K) that
apparently do not have a counterpart in any of the other Galactic GCs observed
so far concerning K abundances (Carretta et al. 2013c).

In the scenario of multiple populations in GCs, Ventura et al. (2012) advocated
that the pattern of abundances observed in NGC~2419 may be explained also by
proton-capture reactions occurring in a temperature range much higher than
usually observed in more normal cluster stars, favoured by the low metallicity
of the cluster.
In these  particular case, the production of Al from destruction of Mg 
would be accompanied by the activation of synthesis of heavier elements, such as
K, Ca, and also Sc by proton-captures on Ar nuclei. The chief signature of this
extreme burning would be the observation of anticorrelations between these
elements and Mg, and in NGC~2419 we verified that the hypothesis by Ventura et
al. well agrees with observations (Carretta et al. 2013c).

NGC~4833 is far from being as massive as NGC~2419, however we observed clear
signature of processing in H-burning at very high  temperature in its abundance
pattern; moreover it is a metal-poor cluster.
We then checked the run of Si, Sc, and Ca as a function of Mg abundances in
NGC~4833, but no anticorrelation was found, apart from that between Si and Mg
already discussed above. There is only a hint of a correlation between Ca and
Sc, that can be intriguing, because formally such a correlation may be expected
if both these elements are produced from burning of Mg under the conditions
invoked by Ventura et al. (2012).
Unfortunately the associated internal errors are large with
respect to the amount of the variations and the correlation is scarcely
statistically significant (the Pearson correlation coefficient is only 0.25;
with a number of degrees of freedom exceeding 70 this implies that the 
correlation is significant only at about 95-98\% in two-tail tests).

Therefore, NGC~4833 cannot be considered a true sibling of NGC~2419, that
still continues to represent an $unicum$ among GCs. Nevertheless, we found in
the present study that NGC~4833, with its associated extreme chemistry, stands 
out among other globular clusters.
To illustrate this finding we used the [Ca/Mg] vs [Ca/H] plane adopted in
Carretta et al. (2013c) as a diagnostic for the relevance of the high
temperature nuclear cycles possibly activated in polluters of the first generation in GCs. While
measurements of the K abundance are still scarce, Ca and Mg are measured for a
large number of stars, providing the precious advantage of good statistics.

\begin{figure}
\centering 
\includegraphics[scale=0.40]{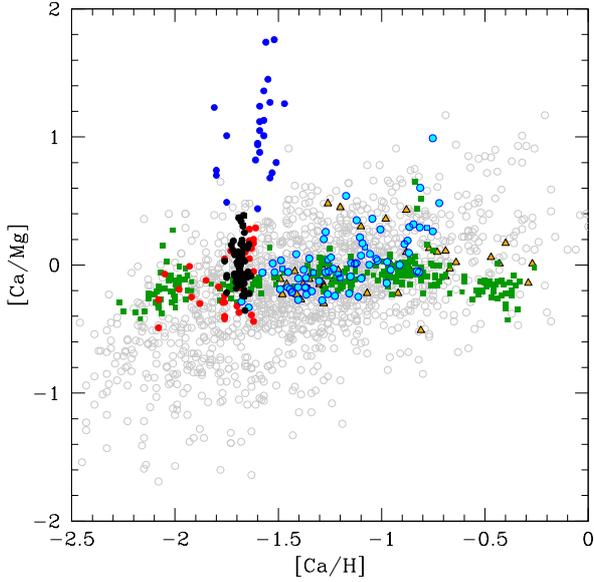}
\caption{The [Ca/Mg] ratio as a function of the Ca abundances for several
stellar populations, adapted from Carretta et al. (2013c). Grey open circles are
stars in eight dwarf spheroidal Milky Way satellite galaxies (Kirby et al. 2011);
blue and red filled circles are RGB  stars in NGC~2419 with [Mg/Fe] lower and
larger than 0.0 dex, respectively,  from Cohen and Kirby (2012) and Mucciarelli
et al. (2012).  Orange triangles are giants in $\omega$ Cen (Norris and Da Costa
1995), cyan circles are RGB stars in M~54 (Carretta et al. 2010b). Green
squares are for giants in 22 Galactic GCs (Carretta et al. 2009b, 2010c, 2011,
2013a,b). Black symbols are stars in NGC~4833 from the present study (squares
and circles for the UVES and GIRAFFE samples, respectively).}
\label{f:versione3}
\end{figure}

In Fig.~\ref{f:versione3} we updated this diagnostic plot by adding stars of
NGC~4833 analyzed in the present paper. Typically, the [Ca/Mg] ratios in GC
stars show small spread over a range of about 2 dex in Ca, with a few
exceptions, represented by few giants in $\omega$ Cen and M~54, the two most
massive clusters in the Galaxy, considered to be the nuclei left from dwarf
galaxies accreted in the past. The three Mg-poor stars in NGC~2808 (Carretta et
al. 2009b) stand out around [Ca/H]$\sim -0.85$ dex, while in the low metallicity
regime the stars of NGC~4833 show a large dispersion in [Ca/Mg] when compared to
other GCs.
The spread in NGC~4833 does not reach the high values of the peculiar Mg-poor
component in NGC~2419, however the stars in NGC~4833 with the largest depletion
in Mg reach the same level of the most Mg-poor stars in $\omega$ Cen. We caution
the reader that among GCs stars small offsets could exist with respect to the
samples by Cohen and Kirby, Mucciarelli et al. and Norris and Da Costa, whereas
all other RGB stars are from the homogeneous analysis by our group.
There is no doubt that NGC~4833 shares some of the peculiarities also seen in 
$\omega$ Cen, M~54 and NGC~2808, although all these GCs are far from reaching
the huge Mg-depletions observed in NGC~2419.

All these objects are high mass clusters, while the extension of the HB and of
the Na-O anticorrelation (and, overall, of the proton-capture processing) are
quite large in NGC~4833 with respect to its absolute magnitude. Hence, we 
could wonder about the reason why NGC~4833 is not so massive, 

In Fig.~\ref{f:iqrmvteff48} we indicated the
position of NGC~288 in the relation between the extension of the Na-O
anticorrelation and total cluster mass (luminosity).
In Carretta et al. (2010a) we discussed the evidence that GCs lying on the left of
this relation lost a larger than average fraction of their mass after their
formation. This conclusion stemmed from old, classical proofs (like tidal tails
associated to NGC~288, Leon et al. 2000) or from new interpretation of
independent observations (like the number density of X-ray sources in M~71, see
Section 5.3 in Carretta et al. 2010a). 

Are there any indication of a huge mass loss also in NGC~4833? The field of view
around NGC~4833 is very crowded (see Fig.~\ref{f:cmdsel4833}), with variable
reddening, and probably this deterred investigations looking for tidal tails
around GCs, because we are not aware of any studies of this kind for NGC~4833.

Another approach could be to look for clues from the present-day mass function
(PDMF). Unfortunately, there seems to be no determination of the PDMF for
NGC~4833, again probably bacause of the difficulties related to the above
mentioned conditions. 
We note that GCs with central brightness $\mu_V$, concentration, and central
density $\rho_0$ similar to those of NGC~4833 present a PDMF with a rather steep
slope of about -1 (de Marchi and Pulone 2007, Paust et al. 2010), indicating 
many low mass stars, and therefore little evidence of mass loss. 

However, these other GCs are characterised by less critical orbits. Dynamical
considerations suggest that NGC~4833 could actually be in a phase of destruction
because of the tidal interaction with the bulge of the Galaxy. 
The concentration of NGC~4833 is modest (c=1.25, Harris 1996) and it is worth 
noting that the cluster has a very eccentric orbit ($e\sim0.84$; Casetti-Dinescu
et al. 2007) passing very close to the Galactic bulge. This is probably a lethal
combination, for a cluster. 

We inserted the data for NGC~4833 (from Harris 1996 and Casetti-Dinescu et al.
2007) in the equation 2 of Dinescu et al. (1999) that estimates the inverse
ratio of the destruction time due to bulge shocking. For NGC~4833 we derived a
destruction time between 1 and 3$\times10^8$ years. The exact value depend on
the mass of the Galactic bulge, assumed to be 34$\times 10^9$ $M_\odot$
(Johnston et al. 1995), and on the velocity at the pericentre of the orbit,
between 200 and 400 km s$^{-1}$. The estimate may increase by a factor 2.5 by
adopting a pericentre distance of 0.9 kpc, at the upper boundary of the range
considered by Casetti-Dinescu et al. (2007), instead of 0.7 kpc. When compared
to the other GCs included in the sample of Dinescu et al. (1999), NGC~4833 show
particularly critical parameters, and therefore it is a good candidate to strong
tidal stripping and destruction by the bulge. Our estimate is approximate, but
these findings are supported by Allen et al. (2006, 2008), who found that
NGC~4833 has the sixth larger destruction rate among the 54 GCs analysed by 
them.

Incidentally, the low value of the $(M/L)_V$ we found (Section 2.1) also
supports  the hint of a significant loss of stars from the cluster, since energy
equipartition should have favored the loss of low mass stars.

\begin{figure}
\centering
\includegraphics[scale=0.40]{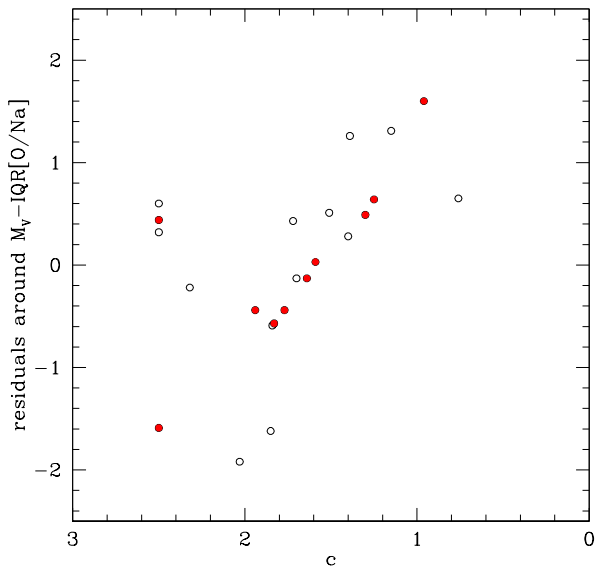}
\caption{Residuals around the relation between total absolute cluster magnitude 
$M_V$ and the interquartile range IQR[O/Na] as a function of the cluster
concentration $c$ for GCs from our FLAMES survey. Filled points indicate inner
halo clusters, as defined in Carretta et al. (2010a).}
\label{f:mviqrc}
\end{figure}

The location of NGC~4833 and NGC~288 on the $M_V - IQR$ diagram suggests that
variations in cluster concentration $c$ can at least in part explain the
observed scatter. In Fig.~\ref{f:mviqrc} we plotted the residuals around the
$M_V - IQR$ correlation against the cluster concentration $c$: there is indeed 
a significant anti-correlation between these two quantities: the Pearson linear
correlation coefficient is $r_p=0.52$ that has a probability lower than 1\% to
be a random result. Of course, in this case it is better to use a bivariate
analysis; the best fit is then:
\begin{equation}
IQR = -0.16 - (0.14\pm 0.04) M_V - (0.16\pm 0.08) c.
\end{equation}
The r.m.s. of residuals around this regression is 0.18, which compares well with
typical internal errors in IQR[O/Na]. The correlation would further improve by
excluding the core-collapse GCs, whose concentration is arbitrarily put at the
constant value 2.5 in Harris (1996). The Pearson linear coefficient provided by
this regression is $r_p=0.65$. This result indicates that at a given absolute
magnitude $M_V$, more loose clusters have a more extended Na-O anticorrelation
than more compact ones. This might seem in contradiction with the observation
that second generation stars are more centrally concentrated than first
generation ones, at least in several clusters (Gratton et al. 2012 and
references therein). However, it may well be explained if the real parameter
driving the Na-O anticorrelation is the original rather than the present day
mass, and more loosely concentrated clusters lost more mass than the more
compact ones. Hence, the original mass of loosely concentrated clusters is on
average larger than that of more compact clusters  that now have the same $M_V$.
More loose clusters are actually expected to be destroyed at a much faster rate
by the tidal effects of the bulge and disk because the destruction rates for
these mechanisms are expected to be proportional to the inverse of the cluster
density (see eqs. (1) and (2) in Dinescu et al. 1999).

Note that if this argument is correct, IQR[O/Na] could be considered a better
proxy to the original cluster  mass than $M_V$. We might then possibly use the
location of a cluster in the $M_V - IQR$ diagram not only to infer its original
mass, but also the amount of mass lost (assuming that very compact clusters
loose only a small fraction of their original mass). This is a quite speculative
but interesting possibility, that should be compared with other dynamical
indicators of mass loss, like e.g. the slope of the mass function. We intend to
make such a comparison in a future study.

\section{Summary}

In the extension of our FLAMES survey of Na-O anticorrelation in GCs 
(e.g. Carretta et al. 2009a,b), we
analyzed FLAMES data for 78 RGB member stars of NGC~4833 (73 observed with
GIRAFFE, 12 with UVES, of which seven in common). This is the first massive
high-resolution spectroscopic study of this cluster.  We confirmed that it is a
metal-poor GCs, with an average [Fe/H]=$-2.04\pm0.003$ from the 73 GIRAFFE
stars and $=-2.015\pm0.004$ from the 12 UVES stars. The $rms$ dispersion is very
small (0.024 and 0.014 dex, respectively), making NGC~4833 one of the most
homogenous GC, at least in its general metallicity.
 
We obtained abundances of Na and O and we concluded that the cluster presents 
the typical Na-O anticorrelation found in (almost) all MW GCs (e.g., Carretta et
al. 2010a, Gratton et al. 2012). The extension of the anticorrelation, as
measured from the interquartile range of the [O/Na] ratio, is large. This is in
line with the expectations based on its very extended HB. The limited sample of
UVES stars shows a marked bimodality in Na and O abundances, as well as in Mg,
Al, Si, with P and I,E stars clearly separated. The GIRAFFE sample shows only a
hint of this separation, because of the larger errors in the abundances.

More exceptional is however the finding of large star-to-star variation in Mg
abundance, anticorrelated with Al, Na, and Si.  The strong depletion in Mg 
implies nuclear processing at very high temeperatures, exceeding a threshold
equal to about 65 million of Kelvin (Arnould et al. 1999). This has been 
found to date only in a few cases, mostly metal-poor and/or massive GCs, with
the more extreme changes seen in objects 
such as the very massive $\omega$ Cen, M~54, and NGC~2419 (and in a
smaller degree in NGC~2808). NGC~2419, with a metallicity similar to NGC~4833,
is a more extreme case and clearly a unicum among GCs in the Milky Way. 
However, NGC~4833 reaches Mg depletion similar to the other massive clusters 
mentioned above and it stands out among GCs of similar low metallicity.

This unusual chemical pattern, coupled with the position of NGC~4833 in the
relation between IQR[O/Na] and mass (total absolute magnitude), seems to 
indicate that the cluster was much more massive in the past. NGC~4833 has
probably lost a conpicuous fraction of its original mass due to bulge shocking,
as also indicated by its orbit.

\begin{acknowledgements}
We thank Chris Sneden for sending us the hyperfine structure components of many
lanthanum lines, and we thank the referee for a careful reading of the
manuscript and suggestions.
This publication makes use of data products from the Two Micron All Sky Survey,
which is a joint project of the University of Massachusetts and the Infrared
Processing and Analysis Center/California Institute of Technology, funded by the
National Aeronautics and Space Administration and the National Science
Foundation.  This research has been funded by PRIN INAF 2011
"Multiple populations in globular clusters: their role in the Galaxy assembly"
(PI E. Carretta), and PRIN MIUR 2010-2011, project ``The Chemical and Dynamical
Evolution of the Milky Way and Local Group Galaxies'' (PI F. Matteucci) . This
research has made use of the SIMBAD database, operated at CDS, Strasbourg,
France and of NASA's Astrophysical Data System.
\end{acknowledgements}

\clearpage

\setcounter{table}{1}
\begin{table*}
\centering
\caption{List and relevant information for target stars in NGC~4833
The complete
Table is available electronically only at CDS.}
\begin{tabular}{rllcccrl}
\hline
    ID    &RA           &Dec          &   $B$ &   $V$ &   $K$  &RV(Hel)&Notes     \\
\hline
 22810  & 12 59  5.559  & -70 56 59.87 &  16.470 &  15.491 &  12.337 &  198.79 & HR11,HR13 \\ 
 23306  & 12 58 49.698  & -70 55 17.19 &  16.736 &  15.696 &  12.439 &  205.33 & HR11,HR13 \\ 
 23437  & 12 58 54.862  & -70 54 59.07 &  16.153 &  15.056 &  11.711 &  195.26 & HR11      \\ 
 23491  & 12 59  5.197  & -70 54 53.86 &  14.242 &  12.516 &   8.270 &  200.33 & HR11,HR13 \\ 
 23518  & 12 59 12.686  & -70 54 50.03 &  16.395 &  15.359 &  12.206 &  203.73 & HR13      \\ 
 24063  & 12 58 54.127  & -70 53 46.86 &  14.660 &  13.343 &   9.651 &  203.13 & HR13      \\ 
 24252  & 12 59 16.083  & -70 53 28.91 &  14.916 &  13.663 &  10.046 &  202.26 & UVES      \\ 
\hline
\end{tabular}
\label{t:coo48}
\end{table*}

\setcounter{table}{5}
\begin{table*}
\centering
\caption[]{Adopted atmospheric parameters and derived iron abundances. The
complete table is available electronically only at CDS.}
\begin{tabular}{rccccrcccrccc}
\hline
Star   &  $T_{\rm eff}$ & $\log$ $g$ & [A/H]  &$v_t$	     & nr & [Fe/H]{\sc i} & $rms$ & nr & [Fe/H{\sc ii} & $rms$ \\
       &     (K)	&  (dex)     & (dex)  &(km s$^{-1}$) &    & (dex)	  &	  &    & (dex)         &       \\
\hline
 22810 & 4893& 2.17 &$-$2.04& 1.46 & 23 & $-$2.037 &0.118&   2 &$-$2.027 &0.115  \\ 
 23306 & 4914& 2.20 &$-$2.03& 1.44 & 21 & $-$2.026 &0.140&   1 &$-$2.021 &       \\ 
 23437 & 4767& 1.89 &$-$2.06& 1.19 &  5 & $-$2.061 &0.062&     &         &       \\ 
 23491 & 4074& 0.45 &$-$2.04& 2.07 & 38 & $-$2.044 &0.105&   3 &$-$2.023 &0.037  \\ 
 23518 & 4867& 2.12 &$-$2.02& 1.43 & 20 & $-$2.017 &0.101&   2 &$-$2.024 &0.001  \\ 
 24063 & 4352& 1.04 &$-$2.05& 1.88 & 26 & $-$2.050 &0.090&   2 &$-$1.985 &0.005  \\ 
 24252 & 4432& 1.20 &$-$1.99& 1.88 & 38 & $-$1.992 &0.058&   5 &$-$1.975 &0.040  \\ 
\hline
\end{tabular}
\label{t:atmpar48}
\end{table*}

\begin{table*}
\centering
\caption[]{Abundances of proton-capture elements in stars of NGC~4833.
n is the number of lines used in the analysis. Upper limits (limO,Al=0)
and detections (=1) for O and Al are flagged.}  
\begin{tabular}{rrccrccrccrcccccc}
\hline
       star  &n &  [O/Fe]&  rms  &  n& [Na/Fe]& rms  &	n& [Mg/Fe]& rms  &  n&[Al/Fe]& rms  &limO&    limAl  & [Na/Fe]$_{\rm LTE}$ & rms\\ %
\hline       
 22810 &  1 &  +0.23 &      &  3 &  +0.53 & 0.04 &  1 &  +0.34 &       &    &        &      &  1 &     &   +0.49&0.05\\ 
 23306 &  1 &  +0.62 &      &  2 &  +0.36 & 0.10 &  2 &  +0.59 & 0.03  &    &        &      &  1 &     &   +0.28&0.11\\ 
 23437 &    &        &      &  2 &  +0.36 & 0.03 &  1 &  +0.39 &       &    &        &      &  1 &     &   +0.22&0.03\\ 
 23491 &  1 &$-$0.01 &      &  3 &  +0.80 & 0.05 &  2 &  +0.25 & 0.02  &    &        &      &  1 &     &   +0.38&0.05\\ 
 23518 &  1 &  +0.55 &      &    &        &      &  1 &  +0.47 &       &    &        &      &  1 &     &        &    \\ 
 24063 &  2 &$-$0.20 & 0.03 &  2 &  +0.89 & 0.00 &    &        &       &    &        &      &  1 &     &   +0.58&0.00\\ 
 24252 &  1 &  +0.48 &      &  2 &  +0.06 & 0.02 &  1 &  +0.58 &       &  1 &  +0.43 &      &  1 &  0  &$-$0.12 &0.02\\ 
\hline
\end{tabular}
\label{t:proton48}
\end{table*}

\begin{table*}
\centering
\caption[]{Abundances of $\alpha$-elements in stars of NGC~4833. 
n is the number of lines used in the analysis.}
\begin{tabular}{rrccrccrccrcc}
\hline
   star      &  n&[Si/Fe]&  rms &    n &  [Ca/Fe]& rms &   n &[Ti/Fe]~{\sc i} &  rms &n &[Ti/Fe]~{\sc ii} & rms \\
\hline   
 22810 &   6  &+0.44 & 0.22 &	4 & +0.36 & 0.05  &  1 & +0.19 &     	&   &       &      \\ 
 23306 &   2  &+0.39 & 0.01 &	6 & +0.35 & 0.09  &  2 & +0.17 & 0.09	&   &       &      \\ 
 23437 &   2  &+0.51 & 0.06 &	1 & +0.31 &       &    &       &     	&   &       &      \\ 
 23491 &   6  &+0.45 & 0.14 &	5 & +0.34 & 0.12  &  4 & +0.15 & 0.05	&   &       &      \\ 
 23518 &   1  &+0.41 &      &	3 & +0.38 & 0.02  &  1 & +0.17 &     	&   &       &      \\ 
 24063 &   1  &+0.40 &      &	5 & +0.37 & 0.05  &  3 & +0.17 & 0.01	&   &       &      \\ 
 24252 &   5  &+0.45 & 0.06 &  13 & +0.35 & 0.05  &  7 & +0.17 & 0.06	& 8 & +0.22 & 0.04 \\ 
\hline
\end{tabular}
\label{t:alpha48}
\end{table*}

\begin{table*}
\centering
\caption[]{Abundances of Fe-peak elements in stars of NGC~4833. 
n is the number of lines used in the analysis.}
\tiny
\setlength{\tabcolsep}{1.2mm}
\begin{tabular}{rrlrlrlrlrlrlrlrlrl}
\hline
      star    & n &[Sc/Fe]~{\sc ii} rms&n& [V/Fe]    rms  &  n &[Cr/Fe]~{\sc i} rms&  n &[Cr/Fe]~{\sc ii} rms&n& [Mn/Fe]   rms  &   n &[Co/Fe]   rms   &  n  &[Ni/Fe]    rms  &  n& [Cu/Fe]  rms&  n &[Zn/Fe]~{\sc i} rms\\
\hline         
 22810 &   4 & -0.04  0.09  &	 &	       &     &       	    &	 &	 	&    &       	    &	 &	       &   3 & -0.16  0.27 &	 &	 	&     &      	   \\  
 23306 &   5 & -0.03  0.24  &	 &	       &     &       	    &	 &	 	&    &       	    &	 &	       &   2 & -0.23  0.19 &	 &	 	&     &      	   \\  
 23437 &   5 & -0.06  0.06  &	 &	       &     &       	    &	 &	 	&    &       	    &	 &	       &   2 & -0.15  0.18 &	 &	 	&     &      	   \\  
 23491 &   6 & +0.00  0.04  &  4 & -0.11  0.03 &   2 & -0.21  0.10  &	 &	 	&    &       	    &  1 & -0.09       &   6 & -0.16  0.11 &	 &	 	&     &      	   \\  
 23518 &   1 & -0.03 	    &	 &	       &     &       	    &	 &	 	&    &       	    &	 &	       &   2 & -0.19  0.06 &	 &	 	&     &      	   \\  
 24063 &   2 & -0.05  0.08  &  2 & -0.09  0.05 &   1 & -0.20 	    &	 &	 	&    &       	    &	 &	       &   3 & -0.16  0.06 &	 &	 	&     &      	   \\  
 24252 &   8 & -0.05  0.14  &  4 & -0.10  0.01 &  12 & -0.21  0.07  &  6 & +0.01  0.08  &  2 & -0.54  0.08  &  1 & -0.05       &   9 & -0.17  0.07 &   1 & -0.85 	&  1  &+0.04 	   \\  
\end{tabular}
\label{t:fegroup48}
\end{table*}

\begin{table*}
\centering
\caption[]{Abundances of $n-$capture elements in stars of NGC~4833 with 
UVES spectra; n is the number of lines used in the analysis.}
\scriptsize
\setlength{\tabcolsep}{1.3mm}
\begin{tabular}{rrccrccrcc}
\hline
  star    & n   &[Y/Fe]~{\sc ii}&rms & n &[La/Fe]~{\sc ii}&rms &n& [Nd/Fe]~{\sc ii} & rms    \\
\hline         
24081	  & 10  &$-$0.15  & 0.11  &    4 & +0.30  & 0.16 &  4 &  +0.45 & 0.09	  \\ 
24252	  &  6  &$-$0.24  & 0.07  &    5 & +0.35  & 0.12 &  4 &  +0.47 & 0.05	  \\ 
31163	  & 10  &$-$0.11  & 0.08  &    3 & +0.25  & 0.07 &  4 &  +0.36 & 0.10	  \\ 
31332	  & 10  &$-$0.09  & 0.10  &    4 & +0.40  & 0.13 &  4 &  +0.39 & 0.11	  \\ 
33347	  &  9  &$-$0.19  & 0.09  &    3 & +0.30  & 0.04 &  4 &  +0.44 & 0.10	  \\ 
33554	  & 10  &$-$0.13  & 0.09  &    4 & +0.25  & 0.15 &  4 &  +0.42 & 0.09	  \\ 
35680	  &  5  &$-$0.09  & 0.06  &    3 & +0.35  & 0.06 &  4 &  +0.40 & 0.11	  \\ 
36402	  &  7  &$-$0.24  & 0.09  &    2 & +0.25  & 0.22 &  3 &  +0.44 & 0.04	  \\ 
36454	  &  9  &$-$0.16  & 0.13  &    3 & +0.25  & 0.11 &  3 &  +0.36 & 0.07	  \\ 
36484	  &  9  &$-$0.12  & 0.09  &    3 & +0.40  & 0.06 &  4 &  +0.41 & 0.13	  \\ 
37197	  &  9  &$-$0.23  & 0.14  &    2 & +0.50  & 0.05 &  3 &  +0.40 & 0.15	  \\ 
37498	  &  9  &$-$0.10  & 0.10  &    4 & +0.25  & 0.07 &  4 &  +0.46 & 0.07	  \\ 
\hline
\end{tabular}
\label{t:neutron48}
\end{table*}

\begin{table*}
\centering
\caption[]{Abundances of Ba~{\sc ii} in stars of NGC~4833. 
n is the number of lines used in the analysis.}
\begin{tabular}{rrcc}
\hline
   star  &  n &  [Ba/Fe]{\sc ii} & rms  \\
\hline   
 22810  &  1 &$-$0.10 &      \\  
 23306  &  1 &$-$0.50 &      \\  
 23437  &    &        &      \\  
 23491  &  1 &$-$0.07 &      \\  
 23518  &  1 &$-$0.33 &      \\  
 24063  &  1 &$-$0.16 &      \\  
 24252  &  3 &$-$0.13 & 0.01 \\  
\hline
\end{tabular}
\label{t:ba48}
\end{table*}

\end{document}